\documentclass[referee]{raa}            

\usepackage{graphicx,times}             
\usepackage{aas_macros} 
\usepackage{natbib}

\bibpunct{(}{)}{;}{a}{}{,}

\usepackage{amssymb}
\usepackage{savesym}
\usepackage{amsmath}
\savesymbol{iint}
\usepackage{txfonts}
\restoresymbol{TXF}{iint}

\usepackage[utf8]{inputenc}
\usepackage{natbib}
\usepackage{mathrsfs}
\usepackage{ulem}
\providecommand{\Gaia}{{\it Gaia} }
\providecommand{\GaiaHVS}{{\it{Gaia}}-HVSC}
\providecommand{\GaiaVS}{HV}
\providecommand{\Unit[1]}{\ensuremath{\mathrm{~#1}}} 

\begin{document}

\title{Three New Late-type Hypervelocity Star Candidates from \Gaia DR2 by Refined Selection Criteria}

\author{Jiao~Li\inst{1,2,3,4}
        \and Shi~Jia\inst{5,1,2}
        \and Yan~Gao\inst{1,2,3,4}
        \and Dengkai~Jiang\inst{1,2,4}
        \and Thomas~Kupfer\inst{6,7}
        \and Ulrich~Heber\inst{8}
        \and Chao~Liu\inst{9}
        \and Xuefei~Chen\inst{1,2,4}
        \and Zhanwen~Han\inst{1,2,3,4}
        }
\institute{Yunnan observatories, Chinese Academy of Sciences, 
           P.O. Box 110, Kunming 650011, China\\
           \email{lijiao@ynao.ac.cn; sjia@must.edu.mo; zhanwenhan@ynao.ac.cn}
           \and
           Key Laboratory for the Structure and Evolution of Celestial Objects, 
           Yunnan observatories, Chinese Academy of Sciences
           \and
           University of Chinese Academy of Sciences, Beijing 100049, China
           \and
           Center for Astronomical Mega-Science, Chinese Academy of Sciences, Beijing 100012, China
           \and
           State Key Laboratory of Lunar and Planetary Sciences, Macau University of Science and Technology, Macau, China
           \and
           Kavli Institute for Theoretical Physics, University of California, Santa Barbara, CA 93106, USA
           \and
           Department of Physics, University of California, Santa Barbara, CA 93106, USA
           \and
           Dr. Remeis-Sternwarte \& ECAP, University of Erlangen-N\"urnberg, Erlangen, Germany
           \and
           Key Lab of Optical Astronomy, National Astronomical Observatories, Chinese Academy of Sciences, Beijing 100012, China
          }

\abstract
{
Several dozen hypervelocity star (HVS) candidates have been reported based on the second data release of \Gaia (\Gaia DR2). However, it has been proven that the radial velocities of some \Gaia HVS candidates are not reliable. In this paper, we employ refined astrometric criteria to re-examine \Gaia DR2, arriving at a more reliable sample of HVS and high velocity star candidates than those found by previous authors. 
We develop a method called Binary Escape Probability Analysis to identify some HVS candidates. This method allows us to work with stars having only two epochs of measured radial velocity. These stars were usually discarded in previous similar studies.
A scrutiny of our final results sheds light on selection effects present in our studies, which we propose to be the focus of future studies. In total, we find three late-type (2 G-type and 1 K-type) HVS and 21 high velocity star candidates, 3 and 11 of which are new, respectively. Judging by their historical trajectories, which we calculate, all three HVS candidates could not have had Galactic centre origins. Further monitoring is required to confirm their status.
}


\maketitle

\section{INTRODUCTION}
\label{introduction}

Hyper-velocity stars (HVSs)\footnote{HVSs are generally thought to be the stars ejected by the Galaxy’s central massive black hole at speeds that can potentially unbind them from the Galaxy. In our paper, we define all unbound stars as HVSs, for the sake of simplicity.} are important tools for probing the Galactic structure (e.g., \citealt{Kenyon2008, Luyoujun2010, Kenyon2014, Brown2015}). HVSs, which were first theoretically predicted by \cite{Hills1988}, are usually defined as stars which can escape the gravitational potential of the Milky Way (MW). Since the first HVS was discovered in 2005 (\citealt{Brown2005}), some further HVS candidates have been found in recent years (e.g., \citealt{Edelmann2005}; \citealt{Hirsch2005}; \citealt{Brown2006};\citealt{Brown2009, Brown2012, Brown2014}; \citealt{Tillich2009}; \citealt{Li2012}; \citealt{Palladino2014}; \citealt{Zheng2014}; \citealt{Zhong2014}; \citealt{Geier2015,Li2015}; \citealt{Huang2017}). However, several of the late type HVS candidates have been rejected from ground based astrometry \citep{Ziegerer2015}. These candidates cover a wide range of spectral types from OBA stars to FGK stars (see the open fast stars catalog from \citealt{Boubert2018}). However, the origin of HVSs is still unclear. According to current understanding (see the review \citealt{Brown2015}), HVSs may have either been formed in the Galaxy or an extragalactic source.

The high velocity of HVSs can be attributed to a number of different ejection mechanisms. The mainstream mechanism is the dynamical interactions between stars and the supermassive central black hole of the MW (e.g., \citealt{Hills1988, Yu2003, Zhangfupeng2010}), which corresponds to an origin at the Galactic center (GC). The kinematic properties of S5-HVS1 is consistent with a GC origin (\citealt{Koposo2019}). Alternatively, a fraction of HVSs originate from the Galactic disk (\citealt{Irrgang2018}). These can be produced via supernova explosions in close binary systems (e.g., \citealt{Blaauw1961,Tauris1998,Wang2009,Tauris2015}) or via dynamical ejections in multiple stellar systems (e.g, \citealt{Gvaramadze2009}). For example, HVS2 (also known as US 708) is likely to be the surviving companion star of a helium double-detonation Type Ia supernova (\citealt{Wang2013,Geier2015}). In addition, the hypervelocity white dwarf (HVWD) LP 40-365 \citep{Vennes2017,Raddi2018b,Raddi2018a} and three newly discovered HVWD candidates \citep{Shen2018} may also be related to the surviving companions of Type Ia supernovae. Besides ejection from the MW, HVSs could also originate from disrupted dwarf galaxies (e.g., \citealt{Abadi2009}) or the Large Magellanic Cloud (LMC) (e.g., \citealt{Boubert2016,Boubert2017}). Recently, the B star HE 0437-5439 (HVS3) was found to have a high probability of originating in the LMC (\citealt{Irrgang2018, Erkal2019}) as already proposed by \citet{Edelmann2005}.

So far, the population of confirmed HVSs is dominated by OBA -type stars (\citealt{Brown2015,Erkal2019,Boubert2018}). Estimating the total velocities of these confirmed HVSs relative to the GC has thus far been achieved via their radial velocities alone, due to the difficulties of measuring proper motion precisely. However, a star's tangential velocity (proper motion times distance) can also contribute significantly to its total velocity (\citealt{Palladino2014, Ziegerer2015}).  The European Space Agency satellite \Gaia has made it possible to search for new HVS candidates and to investigate the origins of HVSs with higher-precision proper motion and stellar property measurements (\citealt{Evans2018, Gaia2018, Marchetti2018a}). 

Armed with this new instrument and the knowledge of an observed object's tangential velocity that accompanied it, it was not long before our understanding of HVSs was enhanced. \citet{Brown2018}, \citet{Erkal2019}, and \citet{Irrgang2018} all studied the origins of known HVSs by obtaining the three-dimensional velocities of these objects from their radial and tangential velocities, and extrapolating back in time to see where they came from. Historical archives of possible HVS candidates, generated from a mixture of spectroscopic radial velocities and very crude tangential velocity estimates, were also revisited (\citealt{Boubert2018}). Many of the objects among them, including the hot subdwarfs US 708 (\citealt{Hirsch2005, Geier2015}) and SDSS J013655.91+242546.0 (\citealt{Tillich2009}) as well as LP40-365 (GD 492, \citealt{Vennes2017, Raddi2018a}) and the LAMOST F9 dwarf star Li10 (\citealt{Li2015}) were confirmed to be HVSs. Then \citet{Raddi2019} find other two WD HVS which is similar to LP40-365. Other studies concentrated on systematic searches for new HVS candidates among the \Gaia DR2 data \citep[e.g.][]{Bromley2018,Marchetti2018,Du2019}, and indeed many were found. However, it should be noted here that these new HVS candidates were identified using \Gaia DR2 radial and tangential velocities, the former of which was found to be spurious in some cases due to contamination from neighboring objects (\citealt{Boubert2019}), and the latter of which did not undergo an efficient test of reliability (see below for details). This led to uncertainties among the candidate selection criteria. Further compounding the issue is the fact that different Galactic potential models can also lead to marginally different results as to whether a certain object can escape, given its current position and velocity. The uncertainty introduced by different Galactic potential models can be at least partly remedied by meticulously listing all the high velocity stars, which have not quite achieved hypervelocity status, found by each study, and by cross-referencing these lists with those of other studies. Establishing these lists may have other benefits, as stars travelling at abnormally high velocities are interesting in their own right, even when they have not achieved escape velocity \citep{Capuzzo-Dolcetta2015}. However, even this measure does not obviate the need of refining the selection criteria of HVS candidates, leading to efficient identification of less controversial HVS candidates with as little data wasted as possible, hence this paper.

To improve the identification efficiency of the selection criteria, three measures can be taken. During the initial candidate selection phase, it is wise to ascertain that the proper motion and parallax model fits to \Gaia data are reliable for the sample. We accomplish this by taking advantage of recent studies that have shown that the $RUWE$ (re-normalised unit weight error) statistic (\citealt{Lindegren2018}) is a good indicator of this reliability. The issue of potential binarity among the sample stars can also be more elegantly handled, such that objects which would have enough velocity to escape the Galaxy, even if contamination due to binarity were to be considered, need not be eliminated from the sample. We achieve this by means of BEPA (Binary Escape Probability Analysis), which is an analytical method that we have developed. The details of BEPA will be given later on. Finally, the aforementioned contamination in the spectra used for radial velocity determination can also be addressed, by simply eliminating those stars which suffer from this influence. In this paper, we adopt all three.

In this work, we first select the HVS candidates and high velocity star candidates from \Gaia DR2 catalogues (Section \ref{sec:data}). Initially, we obtain 16 HVS candidates and 23 high velocity star candidates. In Section \ref{sec:BEPA}, we develop the BEPA approach to study the unbound probabilities of our HVS candidates if they were binary systems, especially for the candidates with only two measured radial velocity epochs from \Gaia ($\mathrm{rv\_nb\_transits}$ =2). In Section \ref{sec:spec}, we analyze the reliability of the radial velocities of our candidates. 
Then, we discuss the implications of our work for future studies and the origins of our HVS candidates in Section \ref{sec:dis}.
Finally, we conclude with a summary.

\section{method}
\label{sec:data}

\subsection{The Galactic Space Velocities}
\label{sec:spacev}

\Gaia DR2 contains 1 692 918 784 sources, of which 7 224 631 have median radial velocities and effective temperatures in the range of [3550, 6900] K (\citealt{Katz2018}). 
We select the sources  with parallaxes larger than 5 times parallax errors ($\varpi > 5\sigma_{\varpi}$), the distances of which we directly determine by inverting their parallaxes: $ d = 1/\varpi$ (\citealt{Astraatmadja2016}).  
We assume that the Sun is located on the Galactic disk at z = 0, at a distance of $d_{\odot} = 8.27$ kpc from the GC, that its peculiar velocity relative to the GC is $(U_{\odot}, V_{\odot}, W_{\odot}) = (11.1, 12.24, 7.25)$ km/s, and that the local circular speed of the Sun is $V_{\mathrm{c}} = 238$ km/s (\citealt{Schonich2010, Schonich2012}).
Then, we use the \Gaia astrometric parameters and their associated errors, which we process using TOPCAT\footnote{http://www.star.bris.ac.uk/~mbt/topcat/} (\citealt{Topcat}), to calculate the Galactic rest frame positions and velocities $v_{\mathrm{grf}}$ for the sources with radial velocities.

\begin{figure}
	\centering
	\includegraphics[width=\linewidth, clip]{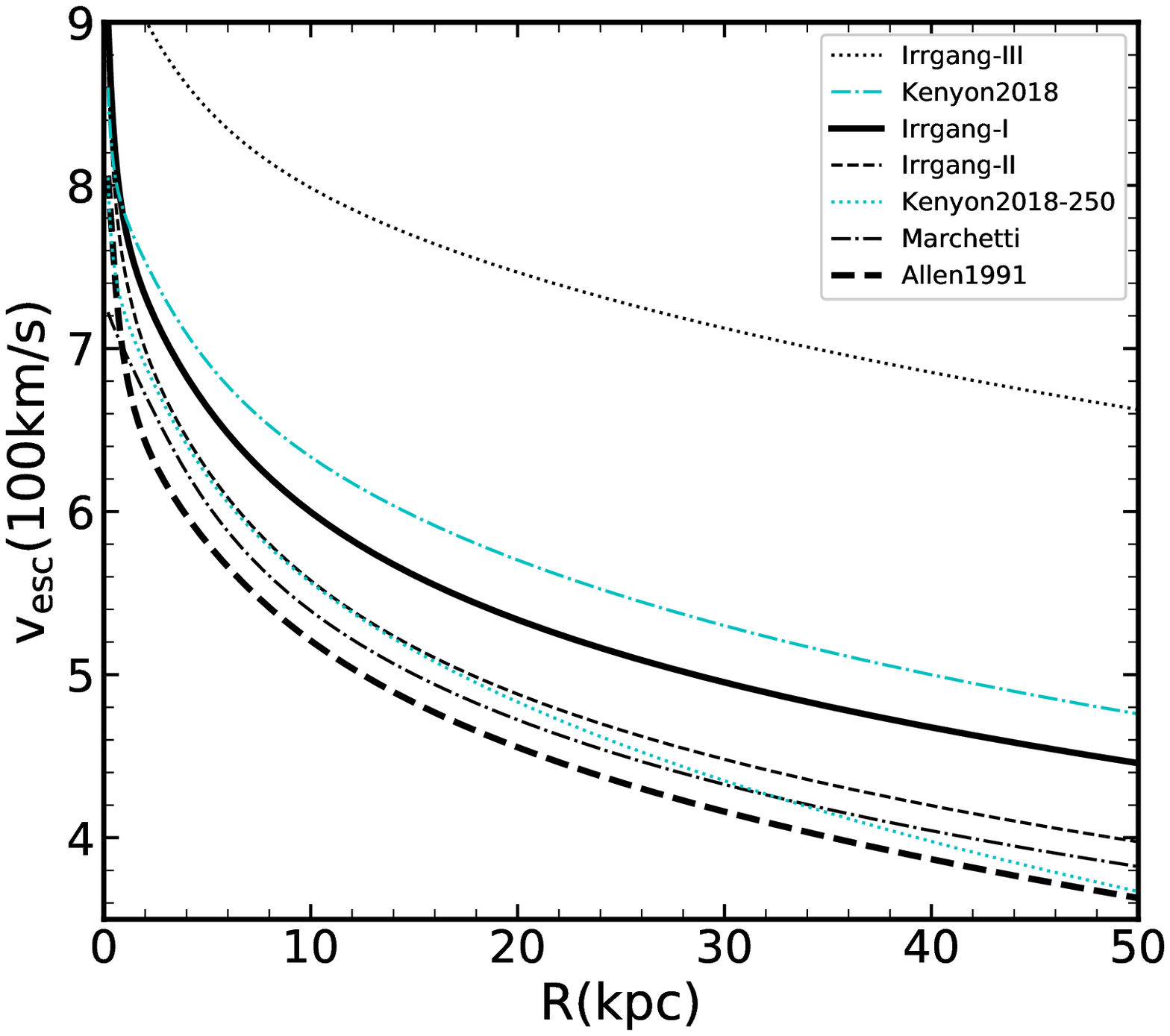}
	\caption{The escape speed curves based on different Galactic potential models at the Galactic disk plane (z=0). Irrgang-I, II, and III are obtained from the potential models of \cite{Allen1991}, \cite{Wilkinson1999} and \cite{Navarro1997}, respectively, and the parameters of these models are updated by \cite{Irrgang2013};  both of Kenyon2018 and Kenyon2018-250  are based on the potential model of \cite{Kenyon2018}, but their escape speeds are calculated by $\sqrt{-2\varphi(\mathbf{r}_{GC})}$ and $\sqrt{2[\varphi(250~\mathrm{kpc}) - \varphi(\mathbf{r}_{GC})]}$, respectively; Marchetti is from the potential model used in \cite{Marchetti2018}; Allen1991 is calculated with the potential model and parameters of \cite{Allen1991}. It should be noted here that the huge escape velocity of Irrang-III is at odds with observational results \citep{Irrgang2018}.
	 }
	\label{fig:galactic}
\end{figure}

\subsection{Selection Criteria}
\label{sec:select}

To filter out data processing artifacts and spurious measurements, we use the following selection criteria:

(a) $\varpi > 5\sigma_{\varpi}$

(b) -0.23 $\leq \rm{mean\_varpi\_factor\_al} \leq$ 0.32

(c) $\rm{visibility\_periods\_used} >$ 8

(d) $\rm{astrometric\_excess\_noise\_sig} \leq$ 2

(e) $\rm{astrometric\_gof\_al} <$  3

(f) phot\_g\_mean\_flux\_over\_error $>$ 20

(g) phot\_bp\_mean\_flux\_over\_error $>$ 20

(h) phot\_rp\_mean\_flux\_over\_error $>$ 20

(i)$\frac{\rm{phot\_bp\_rp\_excess\_factor}}{1.2+0.03(\rm{phot\_bp\_mean\_mag}-\rm{phot\_rp\_mean\_mag})^2} <$ 1.2

(j)$v_\mathrm{grf} > v_\mathrm{esc}$ or $v_{min} > v_\mathrm{esc}$

(k) $RUWE < 1.4$ 

Criteria (a), (b) and (c) ensure that the parallaxes are precise and not vulnerable to errors (\citealt{Astraatmadja2016}).
Criteria (d) and (e) eliminate sources which yield bad astrometric fits (\citealt{gaia_dr2light, Gaia2018}, details can be found in the \Gaia Columns description \footnote{https://gea.esac.esa.int/archive/documentation/GDR2/Gaia\_archive/chap\_datamodel/sec\_dm\_main\_tables/ssec\_dm\_gaia\_source.html}).
Criteria (f), (g), (h) and (i) select sources with good photometry \citep{Evans2018, Eyer2018}, which do not suffer from contamination from nearby sources, and provide relatively good astrometric measurements and radial velocities.
Criterion (k) makes sure that the \Gaia astrometric five-parameter solution is "good". $RUWE$ is the re-normalised unit weight error described by \cite{Lindegren2018} and calculated with the corresponding lookup tables provided at https://www.cosmos.esa.int/web/gaia/dr2-known-issues. It is the equivalent of a reduced ${\chi}^{\rm 2}$ statistic for the five-parameter solution fit. It should be noted that this last criterion was not applied in any previous study in this field to date.

Currently, there are several competing Galactic potential models.
For different models, the escape speed can differ by hundreds of kilometers per second, as shown in Figure~\ref{fig:galactic}.
For our preliminary candidate selection, which should ideally include as many objects as possible, we use the lightest gravitational potential model of \cite{Allen1991} to calculate their escape velocities $v_{\mathrm{esc}}$, and follow up with Potential Model I of \cite{Irrgang2013} later in our paper for more stringent constraints.
After applying criteria (a)-(k), we arrive at 84 candidates.
We then use a Monte Carlo (MC) method to estimate their probabilities of being unbound.

\subsection{Probabilities of being Unbound}
\label{sec:Pun}

We model the coordinate, parallax and proper motion distributions as a multivariate Gaussian distribution with a mean vector $\textbf{m}$ and covariance matrix $\Sigma$ (see Eq. (\ref{eq:mean}), as well as  Eq. (\ref{eq:sigma})):
\begin{equation}
\textbf{m} = (\alpha,  \delta,  \varpi,  \mu_{\alpha},  \mu_{\delta}),
\label{eq:mean}
\end{equation}

\noindent where $\alpha$,  $\delta$,  $\varpi$,  $\mu_{\alpha}$,  $\mu_{\delta}$ are the right ascension, declination, parallax, and proper motions in the direction of the right ascension and declination, respectively;

\begin{equation}
\resizebox{.9 \textwidth}{!} 
{$
\Sigma =
\left(\begin{array}{ccccc}  \sigma_{\alpha}\sigma_{\alpha} & \sigma_{\alpha}\sigma_{\delta}\rho(\alpha, \delta) & \sigma_{\alpha}\sigma_{\varpi}\rho(\alpha, \varpi) & \sigma_{\alpha}
\sigma_{\mu_{\alpha}}\rho(\alpha, \mu_{\alpha}) & \sigma_{\alpha}\sigma_{\mu_{\delta}}\rho(\alpha, \mu_{\delta}) \\
\sigma_{\delta}\sigma_{\alpha}\rho(\delta, \alpha) & \sigma_{\delta}\sigma_{\delta} & \sigma_{\delta}\sigma_{\varpi}\rho(\delta, \sigma_{\varpi}) & \sigma_{delta}\sigma_{\mu_{\alpha}}\rho(\delta, \mu_{\alpha}) & \sigma_{\delta}\sigma_{\mu_{\delta}}\rho(\delta, \mu_{\delta}) \\
\sigma_{\varpi}\sigma_{\alpha}\rho(\varpi, \alpha) & \sigma_{\varpi}\sigma_{\delta}\rho(\varpi, \delta) & \sigma_{\varpi}\sigma_{\varpi}& \sigma_{\varpi}\sigma_{\mu_{\alpha}}\rho(\varpi, \mu_{\alpha}) & \sigma_{\varpi}\sigma_{\mu_{\delta}}\rho(\varpi, \mu_{\delta}) \\
\sigma_{\mu_{\alpha}}\sigma_{\alpha}\rho(\mu_{\alpha}, \alpha) & \sigma_{\mu_{\alpha}}\sigma_{\delta}\rho(\mu_{\alpha}, \delta) & \sigma_{\mu_{\alpha}}\sigma_{\varpi}\rho(\mu_{\alpha}, \varpi) & \sigma_{\mu_{\alpha}}\sigma_{\mu_{\alpha}} & \sigma_{\mu_{\alpha}}\sigma_{\mu_{\delta}}\rho(\mu_{\alpha}, \mu_{\delta}) \\
\sigma_{\mu_{\delta}}\sigma_{\alpha}\rho(\mu_{\delta}, \alpha) & \sigma_{\mu_{\delta}}\sigma_{\delta}\rho(\mu_{\delta}, \delta) & \sigma_{\mu_{\delta}}\sigma_{\varpi}\rho(\mu_{\delta}, \varpi)& \sigma_{\mu_{\delta}}\sigma_{\mu_{\alpha}}\rho(\mu_{\delta}, \mu_{\alpha}) & \sigma_{\mu_{\delta}}\sigma_{\mu_{\delta}}
\end{array} \right)
$}
\label{eq:sigma}
\end{equation}
\noindent where $\sigma_{i}$ is the error of the astrometric parameter $i$ ($i=\alpha,  \delta,  \varpi,  \mu_{\alpha},  \mu_{\delta}$), and $\rho(i,j) = \rho(j,i)$ denotes the correlation coefficients between the astrometric parameters $i$ and $j$, which can be found in the \Gaia DR2 catalog (for example, $\rho(\alpha, \delta)$ is labeled as $\rm{ra\_dec\_corr}$).

We then obtain the Galactic rest frame velocities and estimate the unbound probabilities by combining radial velocities and radial velocity errors:

Radial velocity is measured independently, and hence we assume that it follows a normal distribution, the mean and standard deviation of which are the median radial velocity $v_\mathrm{rad}$ and the radial velocity uncertainty $\epsilon_{v_\mathrm{rad}}$ of \Gaia DR2, respectively.

Devising a MC method, we generate a random position and Galactic rest frame velocity $v_{\mathrm{grf}}$ according to the aforementioned probability distributions for each sample HVS candidate. 
From the random position, we calculate its corresponding  escape velocity $v_{\mathrm{esc}}$, and test whether the HVS candidate is unbound for this particular simulation by comparing $v_{\mathrm{grf}}$ and $v_{\mathrm{esc}}$. 
This process is repeated $10^6$ times, leading to a probability that this HVS candidate is unbound,

\begin{equation}
P_{\mathrm{un}} = \frac{n(v_{\mathrm{grf}} > v_{\mathrm{esc}})}{10^6}
\label{eq:Pun}
\end{equation}
\noindent where $n$ is the number of simulations in which $v_{\mathrm{grf}}>v_{\mathrm{esc}}$.

For the sake of selecting only the sources with reasonable Galactic space velocities, we follow the criterion used by \cite{Marchetti2018}:
\begin{equation}
\sigma_{v_{\mathrm{GC}}}/\tilde{v}_\mathrm{GC} < 0.2
\label{eq:vsigma}
\end{equation}
\noindent where $\tilde{v}_{\mathrm{GC}}$ is the median of $v_{\mathrm{grf}}$ (or $v_\mathrm{min}$) sampled by our previously mentioned MC method, and $\sigma_{v_{\mathrm{GC}}}$ is the square root of the sum of the lower and upper uncertainties on $v_{\mathrm{grf}}$ (see Equation \ref{eq:sigmavGC} )

\begin{equation}
\sigma_{v_{\mathrm{GC}}} = \sqrt{[Per({v_{\mathrm{GC}}, 16) - \tilde{v}_{\mathrm{GC}}]^{2} + [Per(v_{\mathrm{GC}}, 84)-\tilde{v}_{\mathrm{GC}}]^{2}}}
\label{eq:sigmavGC}
\end{equation}
where $Per(v_{\mathrm{GC}}, 16)$ and $Per(v_{\mathrm{GC}}, 84)$ are the 16th and the 84th percentiles of $v_{\mathrm{GC}}$ (or $v_\mathrm{min}$), respectively.

After applying this criterion, our sample consists of 39 candidates (see Table~\ref{tab:rv1} and Table~\ref{tab:hvrv1}).
To obtain our final HVS candidate sample, we repeat the above procedure for the 39 entries, this time using the Galactic Potential Model I of \cite{Irrgang2013}, since this model is the most realistic according to recent studies of the motions of globular clusters and satellite galaxies using \Gaia DR2 astrometry (\citealt{Helmi2018, Watkins2018, Sohn2018, Fritz2018}). We then proceed to demand that $P_\mathrm{un} > 0.8$ for all candidates. This yields 16 HVS candidates (defined as \GaiaHVS) with radial velocities (see Table \ref{tab:rv1} and \ref{tab:rv2}), and the other 23 sources are defined as high velocity star candidates (defined as \GaiaVS, see Table \ref{tab:hvrv1} and \ref{tab:hvrv2}).
It should be noticed that, in our analysis above, 15 of the 16 HVS candidates have had their median radial velocities calculated using only two transits ($\mathrm{rv\_nb\_transits}$ =2). Their apparent radial velocities might be at least partly due to contributions from binary orbits. In the next section, we devise a Binary Escape Probability Analysis (BEPA) approach to derive the probabilities of these HVS candidates being unbound.

\section{Binary Escape Probability Analysis}
\label{sec:BEPA}


For sources with only a few observing plane transits, we cannot be certain whether they are components of binary systems. If they were, then binary orbital velocity can manifest itself in the radial velocity measurements, leading to contamination when we are using these radial velocities to calculate whether or not these objects can become unbound from the Galactic potential. To take this possibility into consideration for our unbound probability calculations, we develop the BEPA method, as detailed below.

 Assuming that a source is a binary star with an orbital eccentricity of zero, then its radial velocity is composed of systemic and orbital velocities.
Its observed median radial velocity can be expressed as
\begin{equation}
\tilde{v}_\mathrm{rad} = v_\mathrm{s} + \tilde{v}^t_\mathrm{b},
\label{eq:vrad}
\end{equation}
\noindent where $v_{s}$ is the systemic radial velocity, and $v^t_\mathrm{b}$ is the projected velocity in the radial direction due to binary orbital rotation.
The radial velocity of the binary is assumed to have a semi-amplitude of $K$,
\begin{equation}
v^t_\mathrm{b} = K\cos(\phi^t),
\label{eq:kcos}
\end{equation}
\noindent where $\phi^t$ is the orbital phase at the observation epoch $t$.
We assume that the radial velocity error is due entirely to orbital motion, in which case the standard deviation of the radial velocity can be expressed as follows,
\begin{equation}
\sigma(v^t_\mathrm{rad}) = \sqrt{\frac{1}{N-1}\sum_{t=1}^{N}(v^t_\mathrm{rad}-\frac{1}{N}\sum_{t=1}^Nv^t_\mathrm{rad})^2} 
                                          = K\sigma_\mathrm{cos(\phi)},
\label{eq:sigmaphi1}
\end{equation}
\noindent where $N$ is the number of observations equal to rv\_nb\_transits. Equation \ref{eq:sigmaphi1} can also be expressed as follows:
\begin{equation}
\sigma_\mathrm{cos(\phi)} =  \sqrt{\frac{1}{N-1}\sum_{t=1}^{N}[\cos(\phi^t)-\frac{1}{N}\sum_{t=1}^N\cos(\phi^t)]^2}.
\label{eq:sigmaphi2}
\end{equation}

By combining Equation (\ref{eq:vrad}), (\ref{eq:kcos}) and (\ref{eq:sigmaphi2}), we find that 
\begin{equation}
v_\mathrm{s} = \tilde{v}_\mathrm{rad}  -  \sigma(v^t_\mathrm{rad})X_{\phi},
\label{eq:vs}
\end{equation}
\begin{equation}
X_{\phi} = \frac{\tilde{\mathrm{cos(\phi)}}}{\sigma_\mathrm{cos(\phi)}}, 
\label{eq:xphi}
\end{equation}
\noindent where $\tilde{\mathrm{cos(\phi)}}$ is the median of $\mathrm{cos(\phi^t)}$.

To escape from the Galaxy, a binary must satisfy the following condition:
\begin{equation}
\left|v_\mathrm{grf}\right| \ge \left|v_\mathrm{esc}\right|.
\label{eq:condition}
\end{equation}

The Galactic rest frame velocity $v_\mathrm{grf}$ of a binary can be obtained via the following relation:
\begin{equation}
v_\mathrm{grf}^2 = av^2_\mathrm{s} + bv_\mathrm{s} +c
\label{eq:relation}
\end{equation}
\noindent where a, b, and c can be calculated using the coordinates, proper motions and parallax by means of a matrix (see Appendix \ref{sec:grtv}).
As shown in Equation~\ref{eq:relation}, $v^2_\mathrm{grf}$  is a quadratic function of systemic radial velocity.

By substituting Equation (\ref{eq:condition}) into ({\ref{eq:relation}}),  the escaping condition can be written as follows:
\begin{equation}
\begin{cases}
v_{s} \ge v_\mathrm{resc1} = \frac{-b+\sqrt{b^2 - 4a(c-v^2_\mathrm{esc})}}{2a}, \\
or \\
v_{s} \le v_\mathrm{resc2} = \frac{-b-\sqrt{b^2 - 4a(c-v^2_\mathrm{esc})}}{2a}.
\end{cases}
\end{equation}
\noindent The minimum $v_\mathrm{grf}$ of the binary is larger than its $v_\mathrm{esc}$ when the relation $b^2 - 4a(c-v^2_\mathrm{esc}) < 0$ is satisfied. In this situation, the binary can always escape the Galaxy.

To calculate the unbound probability of the binary, the escaping condition is expressed with $X_{\phi}$ (Equation \ref{eq:vs}):
\begin{equation}
\begin{cases}
X_{\phi} \le (\tilde{v}_\mathrm{rad}- v_\mathrm{resc1})/\sigma(v^t_\mathrm{rad}), \\
or \\
X_{\phi} \ge (\tilde{v}_\mathrm{rad} - v_\mathrm{resc2})/\sigma(v^t_\mathrm{rad}),
\end{cases}
\label{eq:xphiec}
\end{equation}
\noindent where $\tilde{v}_\mathrm{rad}$ and $\sigma(v^t_\mathrm{r})$ are the median and standard deviation of the radial velocities, respectively. 
According to \cite{Katz2018},
\begin{equation}
\sigma(v^t_\mathrm{rad}) = \sqrt{\frac{2N}{\pi}(\epsilon^2_{v_\mathrm{rad}} -0.11^2)},
\label{eq:sigmav}
\end{equation}
\noindent where $\epsilon_{v_\mathrm{rad}}$ is the radial velocity uncertainty (radial\_velocity\_error).

\begin{figure}
\centering
\includegraphics[width=\linewidth, clip]{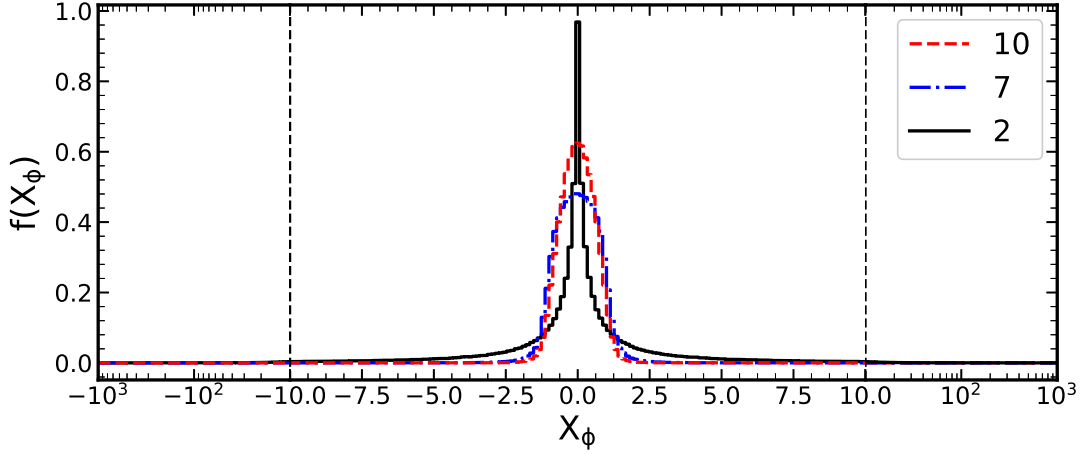}
\caption{The probability density of $X_{\phi}$ for MC simulations with simulated number of observations of 2, 7, and 10, respectively. This provides an estimate of the actual values of $X_{\phi}$ encountered in the \Gaia observations. Note that the horizontal axis is linear within the vertical dashed lines, and logarithmic beyond them.
	 }
\label{fig:f_cosphi}
\end{figure}

Since the orbital phases of the binary system are unknown, we assume that $\phi^{t}$ follows a uniform distribution in the interval [0, $2\pi$] and the binary is observed $N$ times. We use the MC method to generate $N$ random $\phi^{t}$ to compute $X_{\phi}$ with Equation (\ref{eq:sigmaphi2}) and (\ref{eq:xphi}) for each simulation. To obtain the probability density of $f(X_{\phi})$ for a fixed $N$, $10^6$ MC simulations are performed. Figure \ref{fig:f_cosphi} shows an example of the probability density of $X_{\phi}$ with MC simulations for different values of $N$.

\begin{figure}
	\centering
	\includegraphics[width=\linewidth, clip]{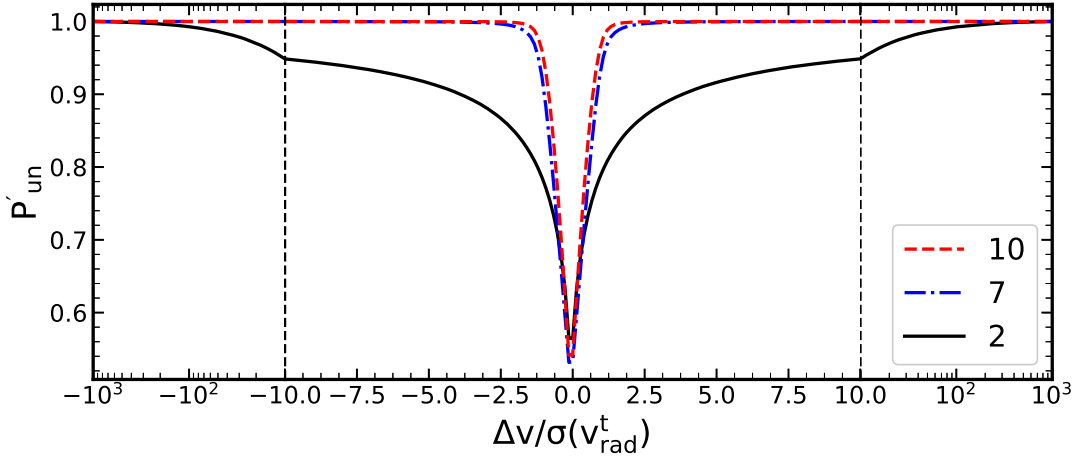}
\caption{The escape probabilities or a series of values of $\frac{\Delta v}{\sigma(v^{t}_\mathrm{rad})}$, where $\Delta v = v_\mathrm{rad}- v_\mathrm{resc, i}$, $(i = 1, 2)$, $v_\mathrm{rad}$ and $\sigma(v^{t}_\mathrm{rad})$ are the median and standard deviation of the radial velocities of \Gaia DR2, respectively. Note that the horizontal axis is linear within the vertical dashed lines, and logarithmic beyond them.
	 }
	\label{fig:Pun_binary}
\end{figure}	
	
According to our approach above, if the source that we investigated is a potential binary system, then its systemic radial escape velocity can be calculated with the median values of $(\alpha, \delta, \varpi, \mu_{\alpha}, \mu_{\delta})$. The probability that the source could escape the Galaxy can then be written as follows:

\begin{equation}
P^{'}_\mathrm{un} =
\begin{cases}
\int_{-\infty}^{(\tilde{v}_\mathrm{rad}- v_\mathrm{resc1})/\sigma(v^t_\mathrm{rad})}\! f(X_{\phi})\, \mathrm{d}X_{\phi}, & (\tilde{v}_\mathrm{rad} \ge v_\mathrm{resc1})\\
                      0, & (v_\mathrm{resc2} \ge \tilde{v}_\mathrm{rad} \le v_\mathrm{resc1}) ~~~~~,\\
\int^{\infty}_{(\tilde{v}_\mathrm{rad}- v_\mathrm{resc2})/\sigma(v^t_\mathrm{rad})} \!f(X_{\phi})\, \mathrm{d}X_{\phi}, & (\tilde{v}_\mathrm{rad} \le v_\mathrm{resc2})
\end{cases}
\end{equation}
\noindent and if $b^2 - 4a(c-v^2_\mathrm{esc}) < 0$, then $P^{'}_\mathrm{un} = 1$.

\begin{figure}
	\centering
	\includegraphics[width=\linewidth, clip]{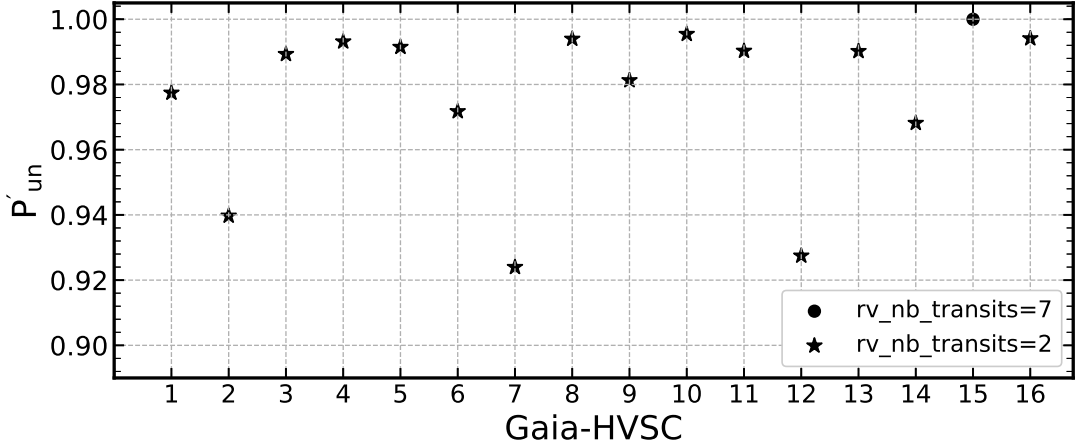}
\caption{ The Binary Escape Probabilities of the 16 sources, which have radial velocities in \Gaia DR2, in the Galactic Potential Model I of \cite{Irrgang2013}. The horizontal number indicates the HVS candidates' number (see Table \ref{tab:rv2}). rv\_nb\_transits is the number of transits (epochs) used to compute the medians and standard deviations of the radial velocities. 
	 }
	\label{fig:BEPA}
\end{figure}

Assuming that our 16 candidates are binary stars, we calculate their binary escape probabilities with the Galactic Potential Model I of \cite{Irrgang2013}. The probabilities are invariably over 92\% for all 16 objects, as shown in Figure \ref{fig:BEPA} and Table \ref{tab:rv2}. This means that if their radial velocity measurements can be assumed to be reliable, then all 16 would almost certainly be hypervelocity objects, whether or not they are in binary systems. However, since incorrect radial velocity measurements have a tendency to manifest themselves as outliers, and hence will be disproportionately represented among stars of high velocity, it would be folly to assume that such extreme cases do not exist in our sample. Consequently, we have no reason to assume that all radial velocity measurements are reliable, which is exactly the issue that we investigate in the next section.
 
\section{Radial Velocities and Possible Contamination}
\label{sec:spec}

As previously noted, it was found by \cite{Boubert2019} that certain spectroscopic radial velocities obtained by \Gaia could be contaminated. They observed  \Gaia DR2 5932173855446728064 (\GaiaHVS15 see Table~\ref{tab:rv1}) at eight epochs with a ground-based telescope, ultimately obtaining a median velocity of $\mathrm{-56.5\pm5.3}$ km/s for the source, which is far slower than the $-614.3\pm 2.5$ km/s found by \Gaia spectroscopy. They point out that, due to the slitless and time delay integration nature of \Gaia, its results are likely to include the light from a close star (at 4.3 arcsec in their case), which is a potential source of contamination. Their studies also find that the radial velocity measurement will be spurious for any star that has a brighter ($G$- or $G_\mathrm{RP}$-band) and closer (less than 6.4 arcsec) neighbor.

To test whether our sample suffers from the same issues, we observe one of our high velocity star candidates (\GaiaVS22, see Table~\ref{tab:hvrv1}) using the Xinglong 2.16m telescope. With a relatively bright magnitude of $G$=13.32 mags, it is impervious to issues arising from low signal to noise ratios (SNR), and has a median radial velocity of $\mathrm{-799.1\pm 1.1}$ km/s and $\mathrm{rv\_nb\_transits}$ = 2, according to the \Gaia catalogue.

\begin{figure}
	\centering
	\includegraphics[width=\linewidth, clip]{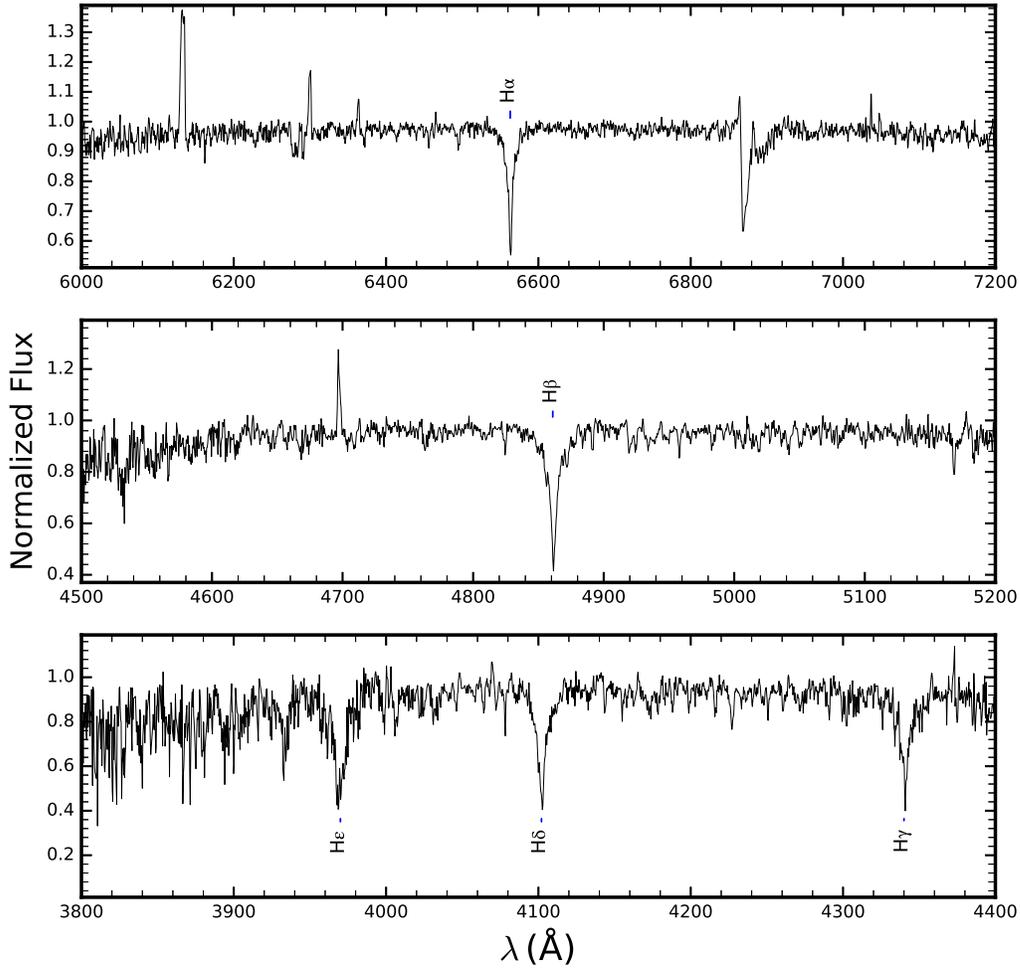}
\caption{ Spectrum of  \GaiaVS22 (\Gaia DR2 1995066395528322560).
	 }
	\label{fig:spec}
\end{figure}
\begin{figure}
	\centering
	\includegraphics[width=\linewidth, clip]{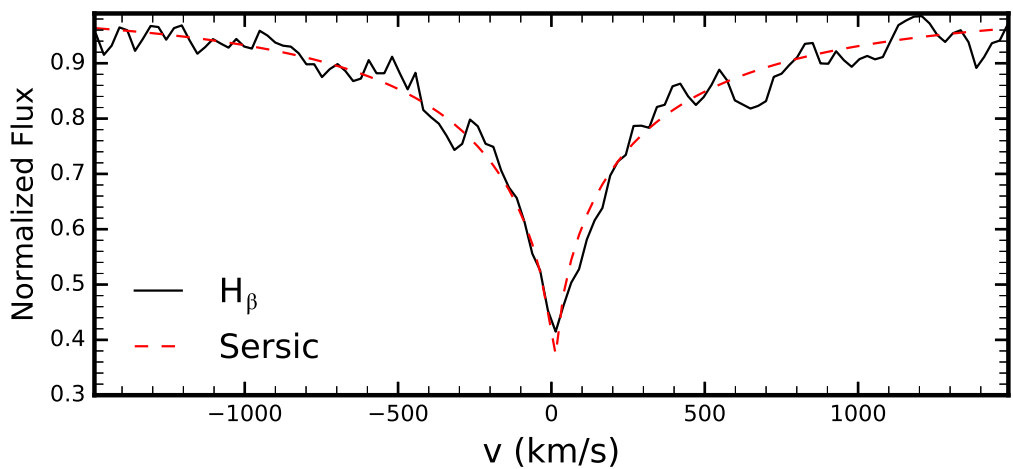}
\caption{$\mathrm{H_\beta}$ line of \GaiaVS22, fitted using a Sersic profile (Equation \ref{sersic}), where $I_{0}$, $v_{0}$, $\sigma$, and $n$ are equal to 0.69,  8.85, 241.11 and 0.59, respectively. The radial velocity derived from this $\mathrm{H_\beta}$ line is 8.85 km/s.
	 }
	\label{fig:sersic}
\end{figure}

\GaiaVS22 was observed on 2019 January 27 using the BFOSC E9+G10 instrument of the Xinglong 2.16-m telescope at Xinglong Observatory (\citealt{Zhao2018}) with a 1.6 arcsec short slit. Its wavelength ranges from 3300 to 10000 $\AA$, and we plot the part of the spectra with a relatively high SNR in Figure~\ref{fig:spec}. Because the Balmer lines have good SNRs, we use Sersic profiles (see Equation~\ref{sersic}) to fit them, and take the velocities corresponding to the centers of the absorption lines to be their radial velocities.

\begin{equation}
f(v) = 1 - I_{o} e^{(\frac{v-v_{0}}{\sigma})^n}
\label{sersic}
\end{equation}
\noindent where $I_{0}$, $v_{0}$, $\sigma$ and $n$ are free parameters for fitting Balmer lines. Physically speaking, $v_{0}$ is the center velocity of an absorption line.
For example, we fit the $\mathrm{H_\beta}$ line with the Sersic profile shown in Figure~\ref{fig:sersic} and obtain a velocity of 8.85 km/s.
Using the same method, we arrive at velocities of 0.08, 23.21, 50.85 and -54.66 km/s for the $\mathrm{H_\alpha}$,  $\mathrm{H_\gamma}$,  $\mathrm{H_\delta}$, and  $\mathrm{H_\epsilon}$ lines, respectively.
We then calculate a radial velocity of $5\pm34$ km/s by using the mean and standard deviation of these 5 velocities from this spectrum.
The radial velocity is much less than the absolute median radial velocity of \Gaia DR2 ($799.1\pm1.1$ km/s, see Table~\ref{tab:hvrv1}). This result confirms the findings of \cite{Boubert2019}, that stars with close neighbors are subject to their spectral contamination, and subsequent spectroscopic radial velocities may not be as reliable as one might hope.

\begin{figure}
	\centering
	\includegraphics[width=0.8\linewidth, clip]{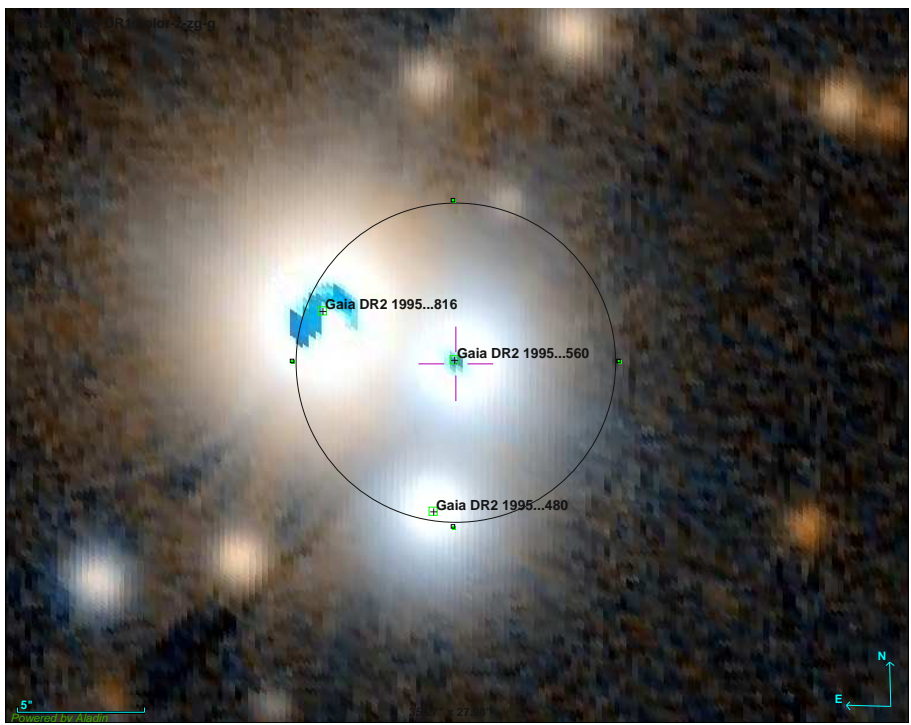}
\caption{PanSTARRS Image centered on GaiaVS22 (\Gaia DR2 1995066395528322560) and dowloaded from Aladin Desktop (http://aladin.u-strasbg.fr). The radius of the circle is 6.4 arcsec. The \Gaia DR2 source IDs of the 3 bright stars in the circle are 1995066395528322816,  1995066395528322560, and 1995066395528324480; their $G$-band magnitudes are 11.85, 13.32, and14.65 mags;  their $G_\mathrm{RP}$-band magnitudes are 11.02, 12.81 and 14.08 mags.
	 }
	\label{fig:panstarrs}
\end{figure}

In Figure~\ref{fig:panstarrs}, we can see that there is a brighter star in the circle centered on \GaiaVS22 with radius of 6.4 arcsec, and another star which is fainter by $\sim1$ mag. Therefore, its \Gaia spectra have a high probability of being polluted by its neighbors, as was expected by \cite{Boubert2019}.

The number of brighter stars around our HVS and high velocity star candidates within 6.4 arcsec
are listed in Table \ref{tab:rv2} and \ref{tab:hvrv2}, respectively.
In Table \ref{tab:hvrv2}, we see that there is another high velocity candidate (\GaiaVS23) having a brighter star within 6.4 arcsec. Its \Gaia radial velocity ($-457.8\pm1.5$ km/s) is therefore unreliable and the Galactic rest frame velocity could hence be totally wrong.
Moreover, the proper motions of these two high velocity candidates are very low, which implies that they are probably not high velocity stars. Finally, 21 high velocity candidates are left with possible ``GOOD" radial velocities, as shown in Table~ \ref{tab:hvrv2}. 
For our 16 HVS candidates,
only five sources do not have brighter companions within 6.4 arcsec.
Of these five remaining sources, we notice that there are significantly bright stars just outside 6.4 arcsec of two of them, \GaiaHVS11 (14.14 mag) and \GaiaHVS12 (14.88 mag), whose \Gaia spectra might also consequently be contaminated. We also eliminate these two objects from our candidate sample for good measure.
Our final candidate sample consists of only \GaiaHVS1, \GaiaHVS2, and \GaiaHVS3, which are consistent with having ``GOOD" \Gaia radial velocities (see Table~ \ref{tab:rv2}).

\section{Discussion}
\label{sec:dis}

In this paper, we employ a set of selection criteria to identify HVS candidates and high velocity candidates from \Gaia DR2 sources which have good photometric and astrometric measurements. With an initial selection, we obtain 16 HVS candidates and 23 high velocity candidates.

Among our 16 HVS candidates, only one candidate (\Gaia DR2 5932173855446728064, \GaiaHVS15 in Table \ref{tab:rv2}) is found amongst the 19 candidates listed by \cite{Marchetti2018}. This is mainly because we use a slightly heavier potential model than theirs. \cite{Marchetti2018} use a four-component Galactic potential model to calculate the escape speed (\citealt{Marchetti2018}), which is lower than the $v_\mathrm{esc}$ obtained from the Galactic Potential Model I of \cite{Irrgang2013}, as shown in Figure \ref{fig:galactic}. Therefore, we find 10 of the 19 HVS candidates listed in \cite{Marchetti2018} to be merely high velocity candidates, instead of hypervelocity ones (see Section \ref{sec:select}, Tables \ref{tab:hvrv1} and \ref{tab:hvrv2}). On the other hand, \cite{Marchetti2018} and \cite{Bromley2018} select candidates with an additional condition $\rm{rv\_nb\_transits} >$ 5, which is not included in our selection criteria. This condition is based on the argument that if a source is just observed a few times ($\rm{rv\_nb\_transits} <$ 5), it is possible that the median radial velocity of the source is caused by either the binary orbit or unreliable \Gaia spectra. With this condition, our 15 HVS candidates are excluded.
To account for any possible impact on our results due to uncertainties in the gravitational potential models, we repeat the process of calculating unbound probabilities for the objects in our sample, this time adding a Gaussian random error with a standard deviation of 30 km/s to the escape velocities. We chose the number 30 km/s because this is the escape velocity difference that one would expect from the gravitational potential models of Irrgang-II and Kenyon2018 depicted in Fig.~\ref{fig:galactic}, at the typical distances (5-12 kpc) from the Galactic centre for our sample objects. The results are also listed in Tables \ref{tab:rv2} and \ref{tab:hvrv2}, where it can be seen that this has little affect on our results.
To investigate the unbound probabilities of the 15 candidates with few radial velocity measurement epochs ($\rm{rv\_nb\_transits} <$ 5), which could potentially be binary components, we develop the BEPA approach. This approach estimates the unbound probabilities of the objects in question under the assumption that they indeed live in binary systems, which we find to be invariably greater than 92\%. Therefore, it is prudent to include them in our HVS candidate sample.

We also note that there is a systemic zero point offset of $\sim -0.067$ mas in the \Gaia parallaxes \citep{Arenou2018}. In order to estimate its influence on our candidates. We calculate the unbound probabilities with distances derived using $1/(\varpi + 0.067\, {\rm mas})$. The unbound probabilities of the HVS candidates are still 1, but the high velocity star candidates are practically no longer able to escape the MW (see Tables~\ref{tab:hvrv1} and \ref{tab:hvrv2}).

However, the BEPA results hinge upon the measurements of the radial velocities, which can, in some cases, be erroneous. For example, 5932173855446728064 (\GaiaHVS15), a HVS candidate from \cite{Marchetti2018},  was found to have an incorrect \Gaia radial velocity determination \citep{Boubert2019}, due to a visible neighbor with similar or greater brightness than the star itself.
We also observed\GaiaVS22 ($G=$ 13.32 mag, $\mathrm{rv\_nb\_transits}$ =2) with the Xinglong 2.16m telescope ourselves, and obtained a radial velocity of $5\pm34$ km/s, which is much less than the median radial velocity of \Gaia DR2 ($-799.1\pm1.1$ km/s). Its \Gaia spectra are likely to be contaminated by its two neighbors within 6.4 arcsec, which is consistent with the result of \cite{Boubert2019}. 
After checking the neighbors of our candidates, only three HVS candidates and 21 high velocity candidates satisfy the condition of not suffering from such spectral contamination.


\begin{figure}[!htbp]
         \centering
         \begin{subfigure}[b]{0.5\textwidth}
              \includegraphics[width=\textwidth]{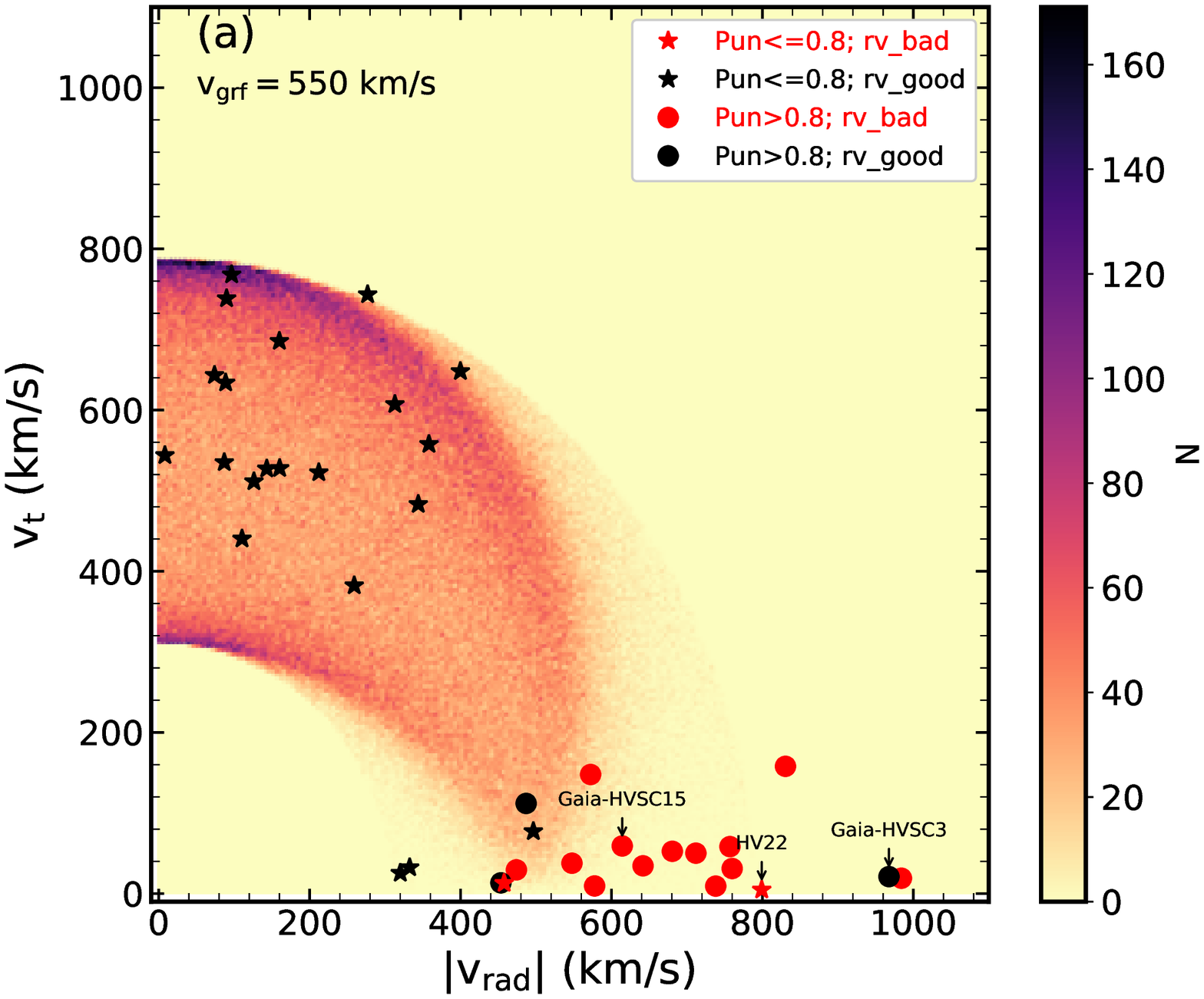}
         \end{subfigure}%
         \label{fig:oaspl_a}
          \begin{subfigure}[b]{0.5\textwidth}
             \includegraphics[width=\textwidth]{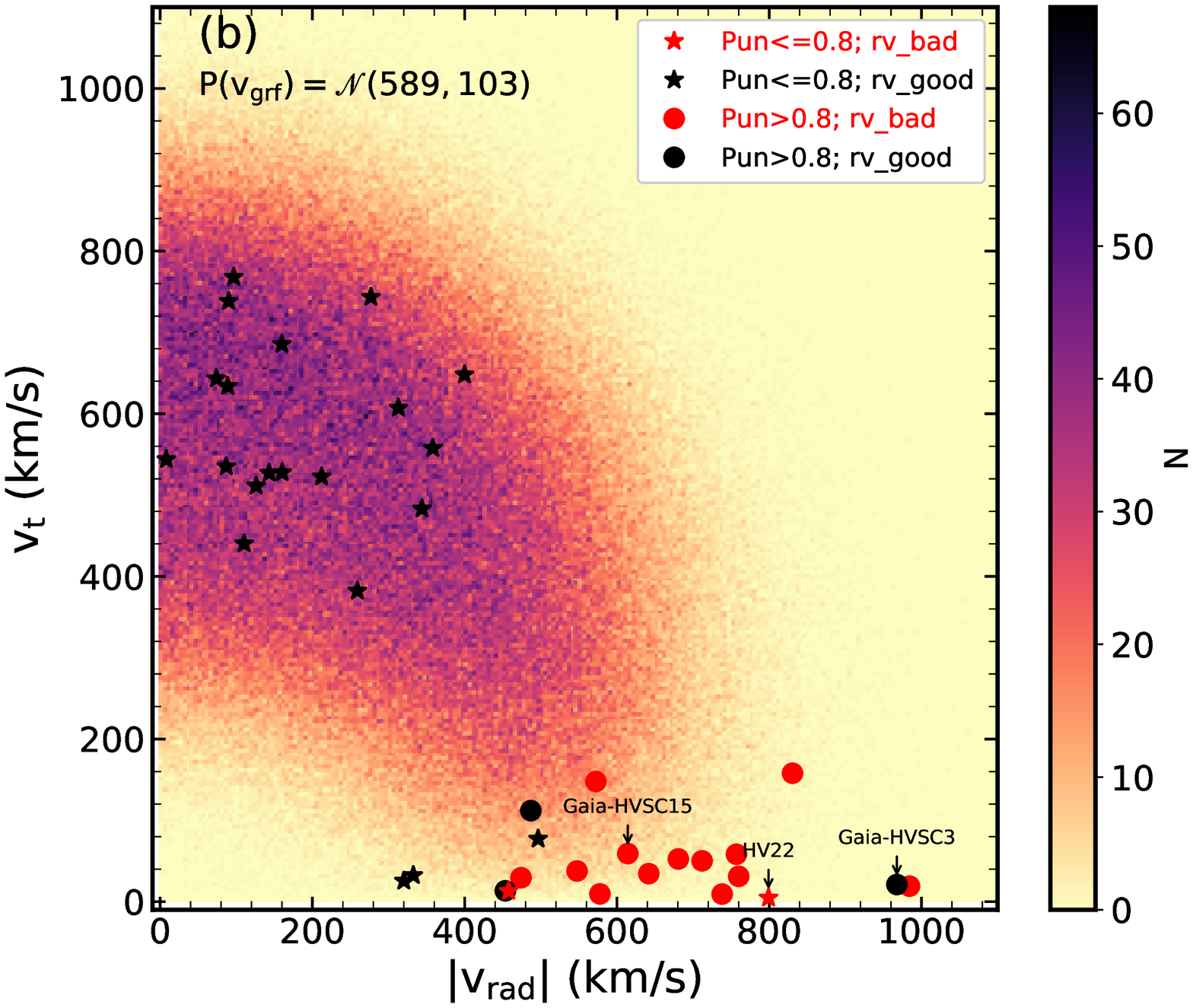}
          \end{subfigure}%
         \begin{subfigure}[b]{0.5\textwidth}
              \includegraphics[width=\textwidth]{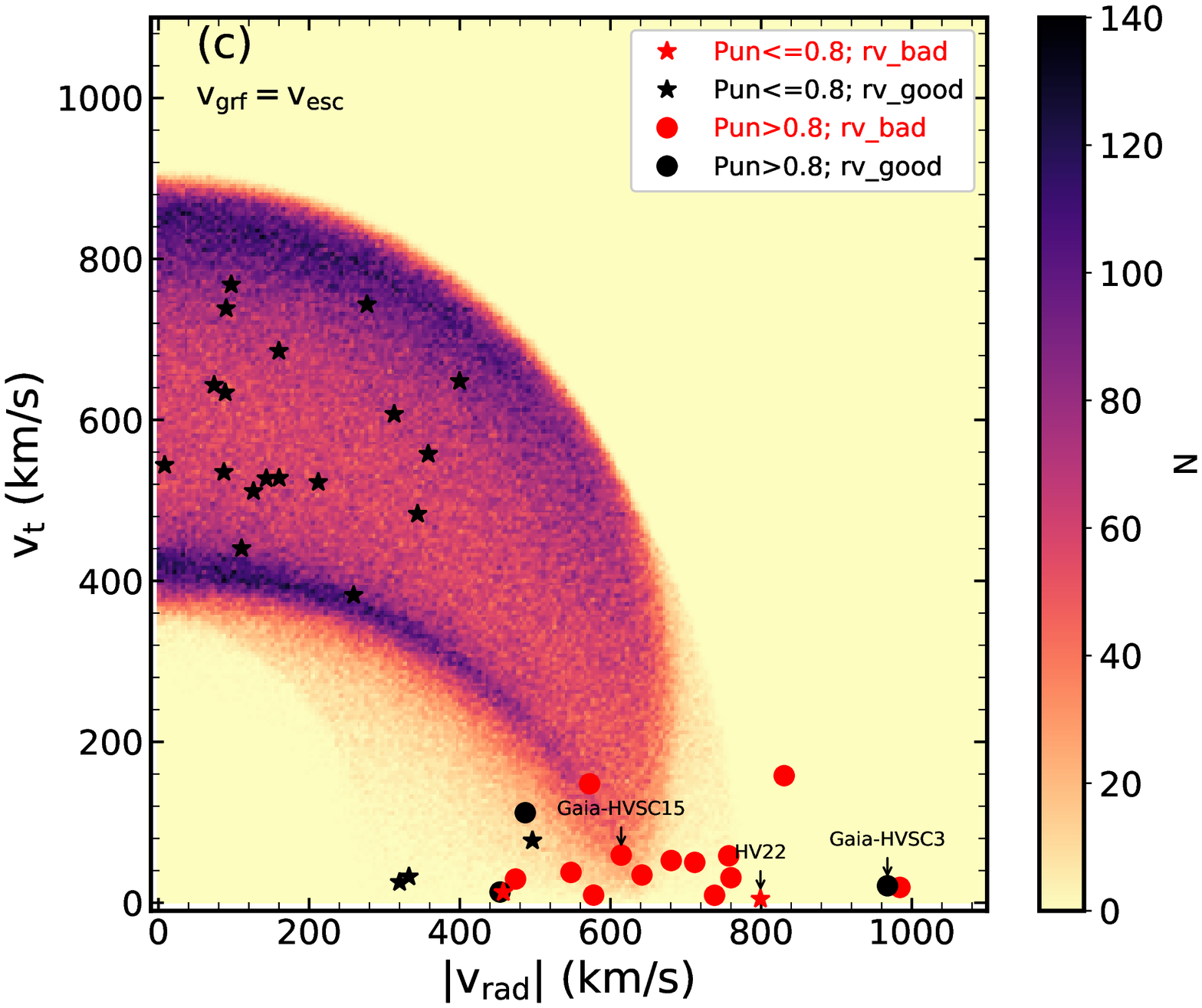}
         \end{subfigure}%
\caption{The radial velocity - transverse velocity diagram. The $y$- and $x$-axes are the tangential and radial velocity moduli, respectively. 
The star-shaped markers are the HVS candidates in Table~\ref{tab:rv1}; the solid circles are the high velocity candidates in Table~\ref{tab:hvrv1};
black and red colors are candidates with ``GOOD" and ``BAD" \Gaia radial velocities, respectively (see Table~\ref{tab:rv2} and \ref{tab:hvrv2});
\GaiaHVS15 and HV22 are the sources observed by \citet{Boubert2019} and us using ground-based telescopes, respectively.
The color bar indicates the number of simulated sources within each velocity bin (details in Section~\ref{sec:summ}). The Galactic rest frame velocities of the simulated sources assumed in each panel are uniformly $v_\mathrm{grf} = 550$ km/s for (a), a Gaussian distribution with a mean of 589 km/s and a standard deviation of 103 km/s for (b), and equal to the local escape velocity of the sources for (c).	 }
	\label{fig:vt_vr}
\end{figure}

\subsection{The radial - transverse velocity Diagram of Candidates}

To visualise our results, we plot our HVS candidates and high velocity star candidates on a radial velocity - transverse velocity plane (see Fig.~\ref{fig:vt_vr}). The objects that were found to have erroneous \Gaia radial velocities are also plotted in red for comparison. In the plot, we can see clearly that most high velocity star candidates lie in areas of high transverse velocity and low radial velocity. Intuitively, this is largely due to velocity directions with higher transverse components taking up a greater solid angle than their high-radial-velocity counterparts. To test that this is indeed the case, we carry out the following experiment.

Noting that most of the candidates have parallaxes larger than 0.14 mas, corresponding to a solar-centric distance of ${\sim}$7 kpc, we artificially generate a mock sample of $10^6$ stars within 7 kpc of the Sun, the number density of which follows that of \citealt{Astraatmadja2016} and references therein. We also stochastically generate the velocities of these objects, assuming a fixed velocity magnitude of 550 km/s (which is the typical velocity of our high velocity star candidate sample), and a spherically random velocity distribution. The distribution of the velocities of these $10^6$ objects relative to the Sun (accounting for solar motion relative to the Galactic Centre) is plotted over our original sample in panel (a) of Fig.~\ref{fig:vt_vr}. It can be seen that the bulk of these simulated objects indeed lie in the region where our high velocity star candidates are to be found. However, it should be noted that some of our sample data points lie beyond this distribution, whereas the lower half $160 \lesssim v_{t} \lesssim 360$~km/s) of this distribution has no data points corresponding to it. Changing the way we generate our mock sample, either by assuming a Gaussian distribution for the velocity  magnitudes (see panel b of Fig.~\ref{fig:vt_vr}), or by setting the velocities to the local escape velocity (see panel c of Fig.~\ref{fig:vt_vr}) does not change this trend. In other words, we do not expect the position of high velocity star candidates within the plane to be due to the previously mentioned solid angle effects alone.

What, then, causes our HVS candidates to lie outside the region covered by our mock sample? What denies the presence of high velocity star candidates in the lower half of the mock sample distribution? The answer is most probably selection effects - it is likely that either the way the \Gaia mission was carried out, or the criteria we use to select our sample, has a tendency to neglect objects that lie in certain regions within this plot. If this interpretation is correct, then the existence of \Gaia-HVSC3 implies the presence of a plethora of HVSs above the region covered by our mock samples in Figs.~\ref{fig:vt_vr}. What the sources of these selection effects may be, however, is beyond the scope of this paper, and will be addressed in future work.

\begin{figure}[!htbp]
         \begin{subfigure}[b]{0.333\textwidth}
              \includegraphics[width=\textwidth]{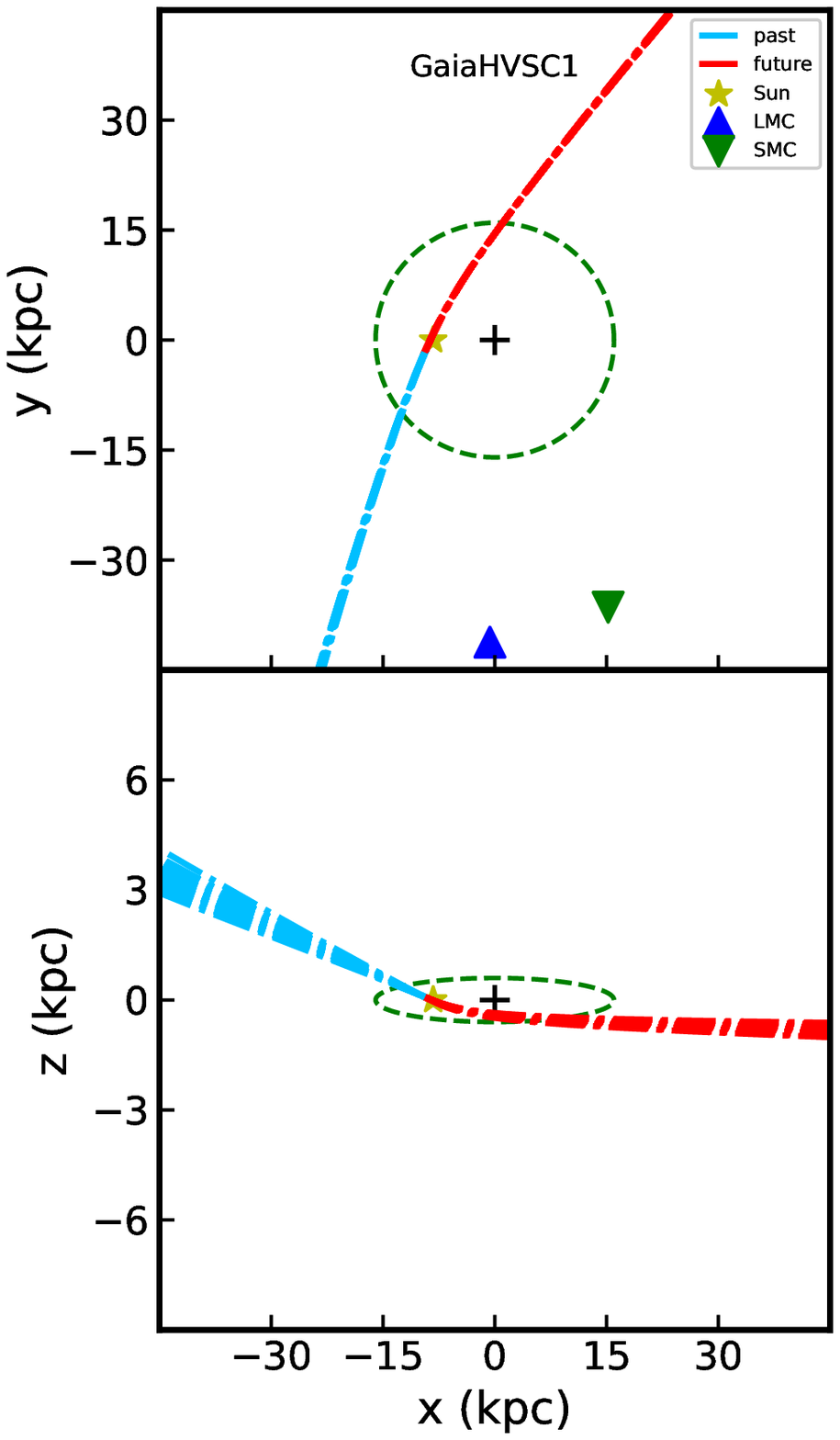}
         \end{subfigure}%
          \begin{subfigure}[b]{0.333\textwidth}
             \includegraphics[width=\textwidth]{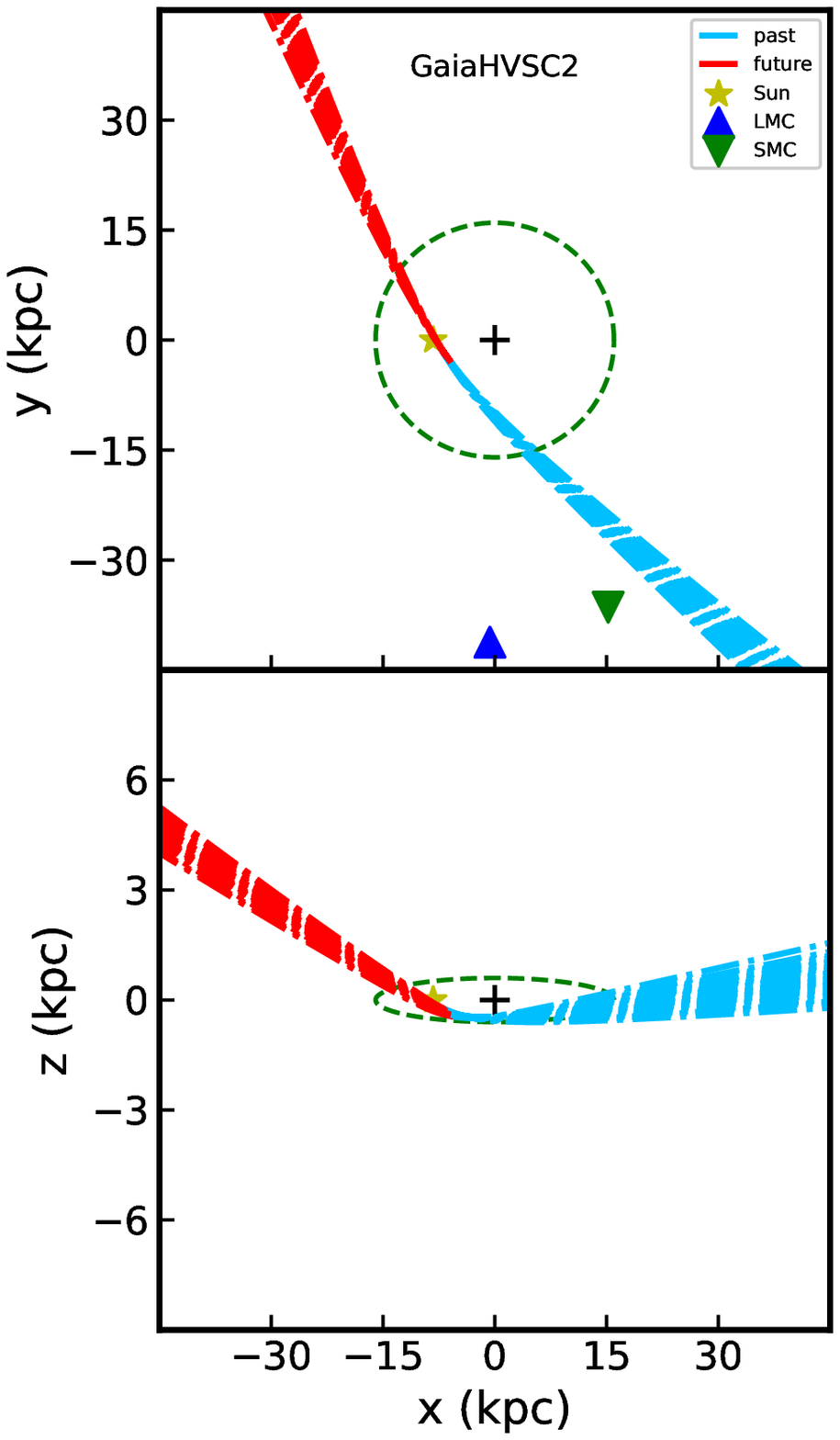}
          \end{subfigure}%
          \begin{subfigure}[b]{0.333\textwidth}
              \includegraphics[width=\textwidth]{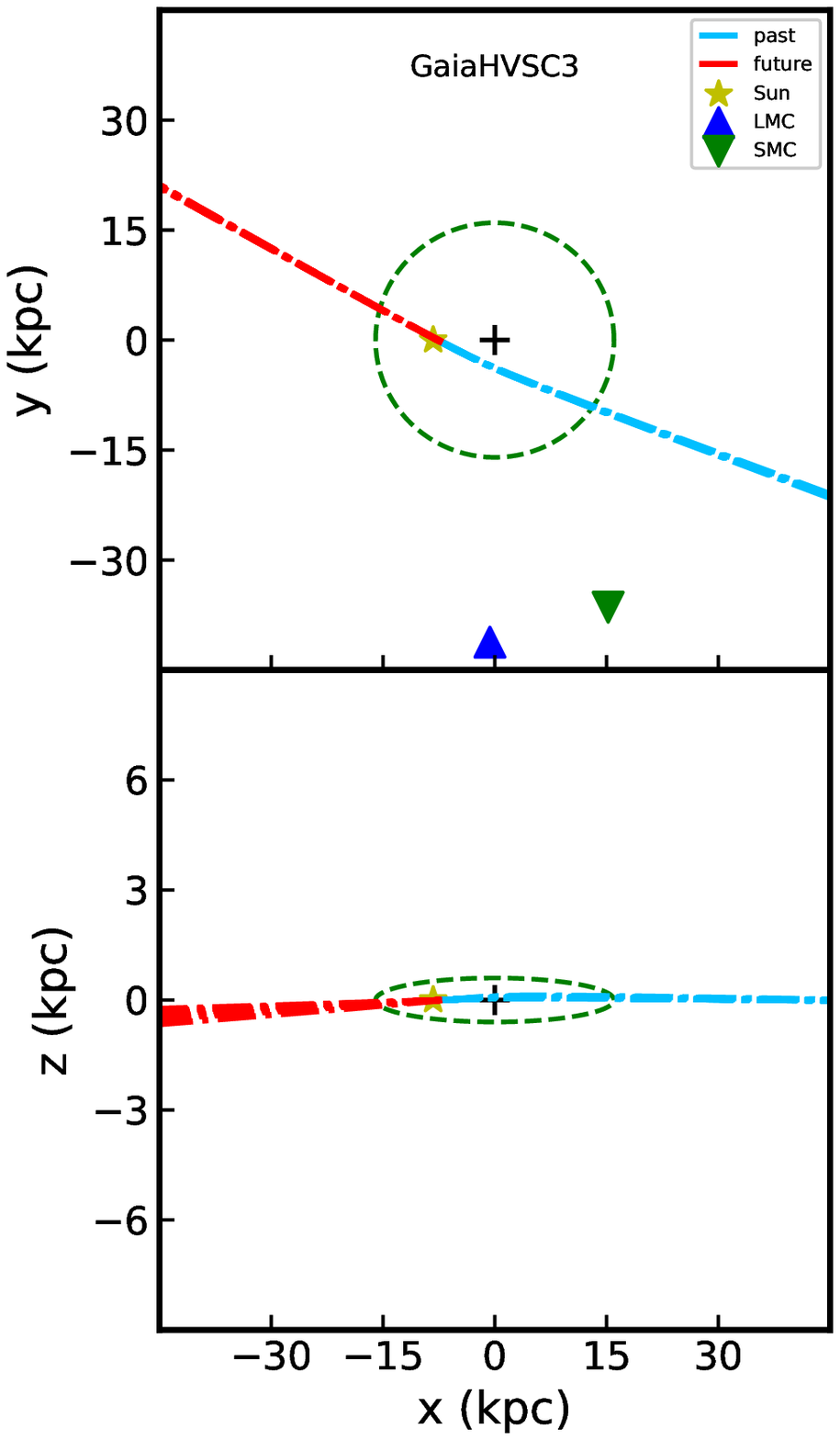}
         \end{subfigure}%
\caption{The integrated past (blue) and future (red) trajectories of the HVS candidates (\GaiaHVS1, \GaiaHVS2 and \GaiaHVS3, see Table~\ref{tab:rv2} ). The blue lines shown are integrated past trajectories, which do not take into account the position of birth of the star; all integrated past trajectories are integrated for 1Gyr, regardless of the age of the star. The star marks the position of the sun, while that of the GC, LMC, and SMC are denoted by a black plus sign, a blue triangle, and a green triangle, respectively. The edge of the Galactic disk at 16 kpc from the GC is marked with a green dashed circle. Both the x-y plane and the x-z plane are plotted for each object in question.
	 }
	\label{fig:trajectories}
\end{figure}

\subsection{HVS Candidate Origins}

To study the origins of HVSs, the simplest way is to trace the positions of our HVS sample back into the past via a set of dynamical calculations, thus obtaining a set of trajectories which shall henceforth be termed integrated past trajectories (IPTs).

Because we do not know when a particular HVS was originated, its past trajectories are integrated over a long timescale to include its birth positions.
Had a HVS only just been born at a point in time $t=t_0$, then its IPT should also include its integrated positions prior to $t_0$.

We calculate the IPTs using the {\it stellar kinematic} code (\citealt{Odenkirchen1992, Pauli2003, Pauli2006}), which calculates trajectories of point masses in the Galactic potential Model I of \cite{Irrgang2013} with a Bulirsch-Stoer integrator.
The trajectories are integrated for 1 Gyr into the past, which we assume to be a generous upper limit to be the time that it would take for an unbound star to escape from the MW. We use a steplength of $\mathrm{d}t = 10^{-4}$ Gyrs (Assuming a HVS with a velocity of 1000 km/s, it will move about 10 pc in every steplength).  

If the unbound probability is less than 100\%, then there exist trajectories which cannot escape the MW and would turn back to the Galaxy after a long travel time.
Since we do not know the ages of the HVS candidates, it is difficult to  determine where they originated from.
This is different, however, for sources which are almost certainly unbound.
From the trajectories, we can easily distinguish the origin of these HVS candidates.

In Figure~\ref{fig:trajectories}, we plot the trajectories of ``Good" HVS candidates. We account for the errors in the \Gaia measurements by running a MC simulation generating the 3-D positions and velocities of these HVS candidates, which take into account the original \Gaia data under the influence of their error bars. These velocities are then used to calculate the IPTs displayed in Figure~\ref{fig:trajectories}, leading to the dispersion of IPTs evident in the figure. The integrated past and future trajectories are indicated by blue and red dash-dotted lines, respectively. 

\GaiaHVS1 moved from the bottom-right to top-left in the x-y plane and has been traveling from the north to the south of MW.
Judging by the fact that it never passed anywhere near the Galactic center, this candidate might either have come from the disk or from the Halo of the MW. For \GaiaHVS2 and \GaiaHVS3, their past trajectories pass closer to the Galactic centre, but not close enough for them to have originated there. From these IPTs, we have no reason to believe that any of these objects are from the Galactic centre.

\begin{figure}
	\centering
	\includegraphics[width=\linewidth, clip]{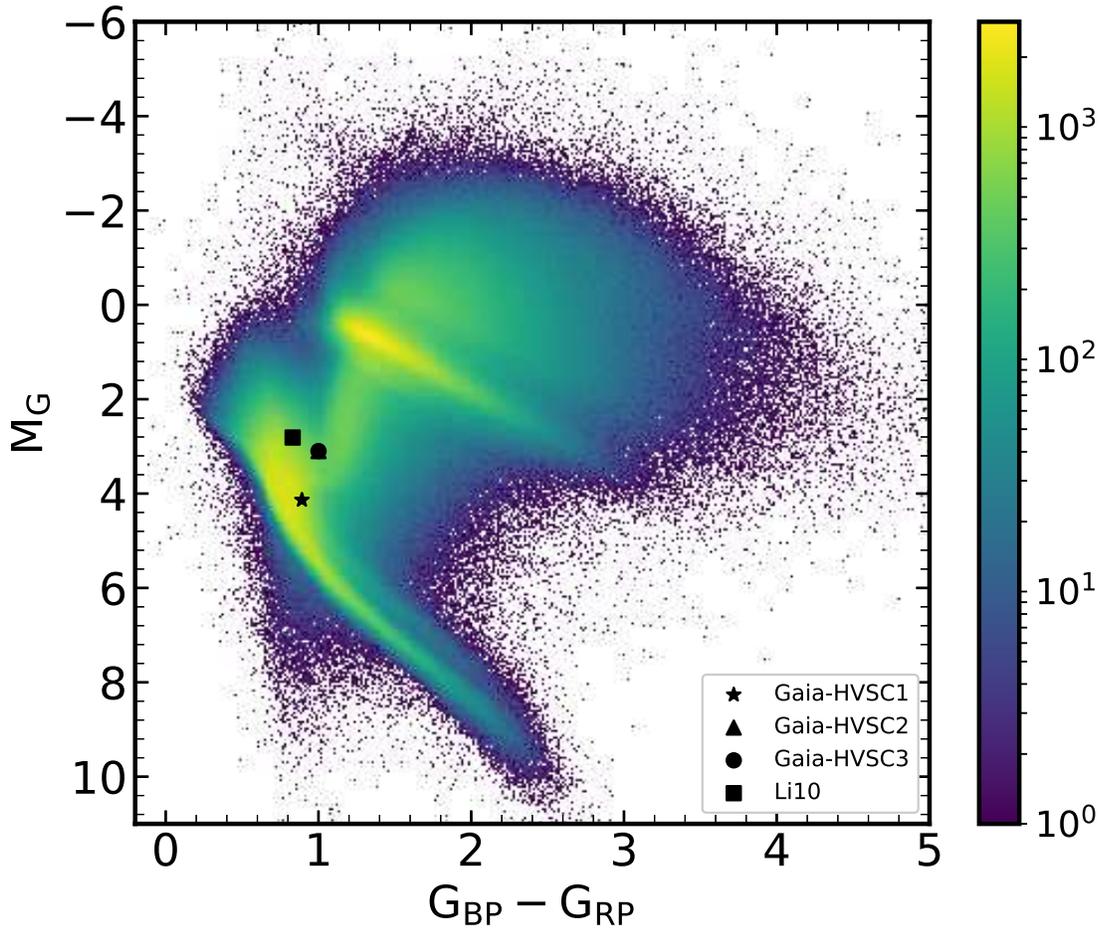}
\caption{ Hertzsprung-Russell (HR) density diagram for about 6.7 million objects which have radial velocities and $\varpi >  5\sigma_{\varpi}$. Black star, triangle and dot are HVS candidates \GaiaHVS1, \GaiaHVS2 and \GaiaHVS3, respectively (see Table~\ref{tab:rv1}). Black square is HVS Li10 \citep{Li2015, Boubert2018}.
	 }
	\label{fig:HR_HVSC}
\end{figure}

Examining the corresponding entries in the \Gaia DR2 catalogue\footnote{https://gea.esac.esa.int/archive/}, we find that \GaiaHVS1, \GaiaHVS2, and \GaiaHVS1 have effective temperatures of about 5629~K (G-type), 5167~K (G-type), and 4166~K (K-type), respectively (see Table~\ref{tab:rv1}). Thus, the 3 HVS candidates are late type stars (see Figure~\ref{fig:HR_HVSC}), similar to HVS Li10 (F-type, \citealt{Li2015, Boubert2018}).  According to traditional wisdom, early (O, B and A) type HVSs are more likely to originate from the Galactic center \citep{Luyoujun2010, Brown2015}, whereas late-type stars can be born in either the Galactic center or the disk. If this is the case, the spectral types of these objects ate consistent with our earlier statement that they did not originate from the Galactic centre, lending further credibility to our conclusions. It should be noted, however, that it has been recently found that a huge fraction of early type HVSs originate from the Galactic disk \citep{Irrgang2018}, therefore the correlation between the origin of an HVS and its spectral type appears to be a weak one. Further data may alter this status quo.

\section{Summary}
\label{sec:summ}

We found three new late-type HVS candidates and 21 high velocity star candidates. Some of our high velocity star candidates are defined as HVS candidates in \cite{Marchetti2018}. 
However, it should be noted that, for some of these new candidates, their $G-$band magnitudes can be close to 15 mags (see Tables \ref{tab:rv1} and \ref{tab:hvrv1}), making them vulnerable to the issues raised in \cite{Katz2018}, namely, that for sources with absolute radial velocities larger than  500 km/s, their radial velocities may be unreliable in the presence of excessively low SNRs. To verify their status as HVSs and high velocity stars, future observations of these objects are necessary. As for the origins of these HVS candidates, we find it unlikely that any of them were born in the Galactic centre.

\begin{table*}[h!]
\caption{HVS candidates with median radial velocities: basic source parameters}
\resizebox{\textwidth}{!}{%
\begin{tabular}{ccc ccc rcccc}
\hline\hline
\GaiaHVS & source id & ($\alpha$, $\delta$) & $\varpi$ & $\mu_{\alpha}$ & $\mu_{\delta}$ & $v_{rad}$ & ($G$, $G_\mathrm{BP}$, $G_\mathrm{RP}$)&$T_{\rm eff}$ &TYPE&$NB_{\rm rv}$ \\
                 &                 &           J2015.5         &   mas     &      mas/yr          & mas/yr              & km/s        &       mag                                                          &  K&\\
\hline
1 & 5716044263405220096 & (115.836451, -19.008715) & $0.52 \pm 0.04$ & $-1.26 \pm 0.05$ & $-0.76 \pm 0.05$ & $-453.5 \pm 2.4$\checkmark&(15.55, 15.91, 15.02)&$5629_{-224}^{+629}$&G&2 \\
2 & 5850309098637075328 & (206.709336, -68.233936) & $0.32 \pm 0.04$ & $-6.68 \pm 0.05$ & $-3.57 \pm 0.05$ & $-486.9 \pm 5.0$\checkmark&(15.57, 15.98, 14.98)&$5167_{-191}^{+187}$&G&2 \\
3 & 5966712023814100736 & (255.893150, -41.563702) & $0.79 \pm 0.07$ & $1.41 \pm 0.13$ & $-3.24 \pm 0.10$ & $-967.7 \pm 5.8$\checkmark&(16.21, 17.08, 15.29)&$4166_{-228}^{+351}$&K&2 \\
\hline
4 & 1825842828672942208 & (296.284240, 20.715550) & $0.75 \pm 0.03$ & $-0.66 \pm 0.04$ & $-5.39 \pm 0.04$ &\textcolor{red}{$641.8 \pm 2.0$}&(14.82, 15.45, 14.03)&$4431_{-57}^{+113}$&K&2 \\
5 & 2251311188142608000 & (301.144379, 70.007552) & $2.89 \pm 0.03$ & $4.60 \pm 0.06$ & $-3.34 \pm 0.07$ &\textcolor{red}{ $738.2 \pm 3.7$}&(15.85, 16.72, 14.87)&$4072_{-198}^{+126}$&K&2 \\
6 & 4065480978657619968 & (273.394905, -24.108792) & $2.34 \pm 0.07$ & $-6.33 \pm 0.10$ & $-25.18 \pm 0.08$ & \textcolor{red}{$-680.7 \pm 1.9$}&(15.47, 16.24, 14.58)&$4159_{-129}^{+144}$&K&2 \\
7 & 4076739732812337536 & (279.020366, -24.132680) & $0.37 \pm 0.03$ & $10.14 \pm 0.05$ & $-5.54 \pm 0.05$ & \textcolor{red}{$572.3 \pm 4.8$}&(13.57, 14.25, 12.76)&$4502_{-124}^{+282}$&K&2 \\
8 & 4103096400926398592 & (278.072328, -15.972720) & $0.64 \pm 0.03$ & $7.71 \pm 0.06$ & $1.25 \pm 0.05$ & \textcolor{red}{$-757.0 \pm 0.7$}&(13.10, 13.79, 12.31)&$4358_{-71}^{+133}$&K&2 \\
9 & 4256598330267724544 & (279.866437, -4.972103) & $0.29 \pm 0.03$ & $1.16 \pm 0.06$ & $-2.03 \pm 0.06$ & \textcolor{red}{$547.6 \pm 1.2$}&(14.34, 15.41, 13.29)&$4007_{-422}^{+981}$&K&2 \\
10 & 4296894160078561280 & (298.560144, 6.421614) & $1.18 \pm 0.05$ & $7.44 \pm 0.07$ & $-2.10 \pm 0.05$ & \textcolor{red}{$760.0 \pm 1.9$}&(15.65, 16.16, 14.94))&$4870_{-45}^{+88}$&K&2 \\
11 & 5305975869928712320 & (146.227409, -57.568968) & $0.29 \pm 0.02$ & $-8.23 \pm 0.04$ & $4.79 \pm 0.05$ & \textcolor{red}{$-830.6 \pm 5.6$}&(14.14, 14.76, 13.19)&$4419_{-138}^{+286}$&K&2 \\
12 & 5412495010218365568 & (145.116991, -45.365443) & $1.15 \pm 0.02$ & $-6.65 \pm 0.04$ & $2.68 \pm 0.04$ & \textcolor{red}{$-474.0 \pm 14.8$}&(14.88, 15.30, 14.29)&$5338_{-163}^{+112}$&G&2 \\
13 & 5878409248569969792 & (217.772803, -61.167859) & $3.06 \pm 0.03$ & $32.36 \pm 0.04$ & $-0.09 \pm 0.06$ & \textcolor{red}{$-711.9 \pm 3.7$}&(12.30, 12.71, 11.73)&$5342_{-35}^{+32}$&G&2 \\
14 & 5931224697615320064 & (249.224053, -51.719940) & $0.45 \pm 0.03$ & $0.50 \pm 0.05$ & $-0.74 \pm 0.03$ & \textcolor{red}{$-577.7 \pm 3.7$}&(13.47, 13.69, 13.03)&$6600_{-322}^{+310}$&F&2 \\
15$^\mathrm{B, M}$ & 5932173855446728064 & (244.118100, -54.440452) & $0.45 \pm 0.03$ & $-2.68 \pm 0.04$ & $-4.99 \pm 0.03$ & \textcolor{red}{$-614.3 \pm 2.5$}&(13.81, 14.21, 13.22)&$5322_{-160}^{+55}$&G&7 \\
16 & 5951114420631264640 & (260.139995, -46.794507) & $0.99 \pm 0.05$ & $2.68 \pm 0.08$ & $2.94 \pm 0.06$ & \textcolor{red}{$-984.3 \pm 3.4$}&(15.50, 15.99, 14.84)&$4938_{-109}^{+224}$&K&2 \\\hline\hline
\end{tabular}}
\label{tab:rv1}
In the first column, the superscript ``B'' and ``M" indicates the sources which are listed in \cite{Bromley2018} and \cite{Marchetti2018} , respectively. $NB_{\rm rv}$ is the number of transits used to compute the medians and standard deviations of the radial velocities (rv\_nb\_transits). $T_{\rm eff}$  is the effective temperature from the Gaia DR2 catalogue. ``TYPE" is the spectral type which is roughly estimated from the corresponding \Gaia DR2 effective temperature. The numbers shown in red are the \Gaia measurements which are known to be erroneous for reasons given in Section~\ref{sec:spec}, shown ``as is", without being corrected to their true physical values. Their unbound probabilities are shown in Table \ref{tab:rv2}.
\end{table*}
\begin{table*}[h!]
\centering
\resizebox{\textwidth}{!}{%
\centering
\begin{tabular}{c ccc | cccc | cccc|ccc}
\hline\hline
                & \multicolumn{3}{c|}{BEPA}                                                                            &            \multicolumn{4}{c|}{$1/\varpi$}                                                                   &  &&&& \\
\hline
\GaiaHVS &     $v_\mathrm{resc1}$   &  $v_\mathrm{resc2}$    &$P^{'}_\mathrm{un}$  &$r_\mathrm{GC}$ & $v_\mathrm{grf}$ & $v_\mathrm{esc}$ & $P_\mathrm{un}$ & $N_\mathrm{RP}$& $N_\mathrm{G}$  &GOOD &  $RUWE$& $v_\mathrm{30, esc}$ & $P_\mathrm{30, un}$&$P_\mathrm{\varpi+0.067, un}$\\
                  &               km/s                &               km/s                  &                                  &          kpc              &   km/s                  &   km/s                     &                             &                              &                               &             &   km/s                     &                             &\\
\hline
1 & 802.59 & -377.88 & 0.978 &$9.485_{-0.098}^{+0.116}$ &$679_{-2}^{+2}$ &$605_{-1}^{+1}$ &1.000 &  0 & 0  &\checkmark &0.947                            &$605_{-30}^{+30}$ &0.993&1.000\\
2 & 822.13 & -441.45 & 0.940 &$6.819_{-0.099}^{+0.094}$ &$680_{-5}^{+5}$ &$635_{-1}^{+1}$ &1.000 &  0 & 0  &\checkmark & 1.056                           &$636_{-30}^{+30}$ &0.931&1.000\\
3 & 642.81 & -532.64 & 0.989 &$7.064_{-0.112}^{+0.095}$ &$1049_{-6}^{+6}$ &$633_{-1}^{+1}$ &1.000 & 0 & 0  &\checkmark & 1.051                           &$633_{-30}^{+30}$ &1.000&1.000\\
\hline
4 & 402.02 & -834.64 & 0.993 &$7.632_{-0.024}^{+0.022}$ &$863_{-2}^{+2}$ &$625_{-0}^{+0}$ &\textcolor{red}{1.000} & 1 & 1 &              & 0.936     &$625_{-30}^{+30}$ &\textcolor{red}{1.000}    &\textcolor{red}{1.000}  \\
5 & 376.52 & -839.47 & 0.991 &$8.349_{-0.001}^{+0.001}$ &$974_{-4}^{+4}$ &$617_{-0}^{+0}$ &\textcolor{red}{1.000} & 1 & 1 &              & 1.033      &$617_{-30}^{+30}$ &\textcolor{red}{1.000}    &\textcolor{red}{1.000}  \\
6 & 549.01 & -633.35 & 0.972 &$7.847_{-0.014}^{+0.013}$ &$668_{-2}^{+2}$ &$623_{-0}^{+0}$ &\textcolor{red}{1.000} & 1 & 1 &              & 1.000       &$623_{-30}^{+30}$ &\textcolor{red}{0.932}    &\textcolor{red}{1.000} \\
7 & 541.64 & -643.60 & 0.924 &$5.660_{-0.214}^{+0.185}$ &$681_{-5}^{+5}$ &$652_{-3}^{+3}$ &\textcolor{red}{1.000} & 1 & 0 &              & 0.929        &$652_{-30}^{+30}$ &\textcolor{red}{0.831}    &\textcolor{red}{1.000}\\
8 & 493.61 & -656.08 & 0.994 &$6.779_{-0.068}^{+0.062}$ &$729_{-1}^{+1}$ &$636_{-1}^{+1}$ &\textcolor{red}{1.000} & 1 & 0 &              & 0.847         &$636_{-30}^{+30}$ &\textcolor{red}{0.999}   &\textcolor{red}{1.000}\\
9 & 501.30 & -749.53 & 0.981 &$5.469_{-0.309}^{+0.257}$ &$700_{-1}^{+1}$ &$656_{-4}^{+5}$ &\textcolor{red}{1.000} & 1 & 1 &              & 0.881          &$656_{-30}^{+30}$ &\textcolor{red}{0.925}  &\textcolor{red}{1.000}\\
10 & 417.29 & -781.86 & 0.995 &$7.715_{-0.022}^{+0.020}$ &$959_{-2}^{+2}$ &$624_{-0}^{+0}$ &\textcolor{red}{1.000} & 2 & 2 &              & 1.174           &$624_{-30}^{+30}$ &\textcolor{red}{1.000} &\textcolor{red}{1.000}\\
11 & 851.75 & -362.16 & 0.990 &$8.406_{-0.064}^{+0.088}$ &$1081_{-6}^{+6}$ &$616_{-1}^{+1}$ &\textcolor{red}{1.000} & 0 & 0  &              &1.202         &$616_{-30}^{+30}$ &\textcolor{red}{1.000} &\textcolor{red}{1.000}\\
12 & 864.58 & -369.26 & 0.928 &$8.294_{-0.001}^{+0.002}$ &$722_{-15}^{+15}$ &$618_{-0}^{+0}$ &\textcolor{red}{1.000} & 0 & 0&              &0.948         &$618_{-30}^{+30}$ &\textcolor{red}{0.999}&\textcolor{red}{1.000}\\
13 & 746.39 & -405.76 & 0.990 &$8.043_{-0.002}^{+0.002}$ &$912_{-4}^{+4}$ &$620_{-0}^{+0}$ &\textcolor{red}{1.000} &  1 & 1 &              & 1.012       &$620_{-30}^{+30}$ &\textcolor{red}{1.000}&\textcolor{red}{1.000}\\
14 & 701.16 & -501.06 & 0.969 &$6.331_{-0.125}^{+0.111}$ &$715_{-3}^{+3}$ &$643_{-2}^{+2}$ &\textcolor{red}{1.000} &  1 & 1&              & 1.153        &$643_{-30}^{+30}$ &\textcolor{red}{0.991}&\textcolor{red}{1.000} \\
15 & 735.08 & -502.80 & 1.000 &$6.461_{-0.113}^{+0.101}$ &$749_{-3}^{+3}$ &$641_{-1}^{+2}$ &\textcolor{red}{1.000} &  0 & 1&              & 1.000         &$641_{-30}^{+30}$ &\textcolor{red}{1.000}&\textcolor{red}{1.000}\\
16 & 639.01 & -506.58 & 0.994 &$7.317_{-0.052}^{+0.047}$ &$1082_{-3}^{+3}$ &$629_{-1}^{+1}$ &\textcolor{red}{1.000} & 0 & 1 &              &1.044         &$629_{-30}^{+30}$ &\textcolor{red}{1.000}&\textcolor{red}{1.000}\\
\hline\hline
\end{tabular}}
\caption{Unbound probabilities of HVS candidates with median radial velocities.
$v_\mathrm{resc1}$ and $v_\mathrm{resc2}$ are escape velocities in the radial direction (see Section \ref{sec:BEPA});
$P^{'}_\mathrm{un}$ is the binary escape probability derived by BEPA, i.e. assuming that the source is a binary system;
$r_\mathrm{GC}$ is the distance to the Galactic center; 
$v_\mathrm{grf}$ is the Galactic rest frame velocity; 
$v_\mathrm{esc}$ is the escape velocity of the Galactic Potential Model I of \cite{Irrgang2013}; 
$P_\mathrm{un}$ is the unbound probability, if its median radial velocity were the systemic radial velocity;
$N_\mathrm{RP}$ is the number of stars brighter than the object in question in the $G_\mathrm{RP}-$band within 6.4 arcsec;
$N_\mathrm{G}$ is the number of stars brighter than the object in question in the $G-$band within 6.4 arcsec;
``GOOD" stands for the candidates with possibly trustworthy radial velocities (detail see Section~\ref{sec:spec});
$RUWE$ is the re-normalised unit weight error,  for which when $RUWE <$ 1.4, it indicates a ``good" solution for astrometric five-parameter fit (\citealt{Lindegren2018}, https://www.cosmos.esa.int/web/gaia/dr2-known-issues).
The numbers shown in red are the Gaia measurements which are known to be erroneous for reasons given in Section 4, shown “as is”, without being corrected to their true physical values.
$P_\mathrm{30, un}$ is the unbound probability calculated with $v_\mathrm{30, esc}$, which is $v_\mathrm{esc}$ with a Gaussian random error of 30 km/s added to it.
$P_\mathrm{\varpi+0.067, un}$ is the unbound probability calculated with distance derived by $1/(\varpi+0.067\, {\rm mas})$, which considers a \Gaia parallax offset of $\sim -0.067$ mas \citep{Arenou2018}.
}
\label{tab:rv2}
\end{table*}
\begin{table*}[h!]
\caption{High velocity star candidates with median radial velocities: basic source parameters}
\resizebox{\textwidth}{!}{%
\begin{tabular}{ccc ccc rcccc}
\hline\hline
\GaiaHVS & source id & ($\alpha$, $\delta$) & $\varpi$ & $\mu_{\alpha}$ & $\mu_{\delta}$ & $v_{rad}$ & ($G$, $G_\mathrm{BP}$, $G_\mathrm{RP}$)&$T_{\rm eff}$ &TYPE&$NB_{\rm rv}$ \\
                 &                 &           J2015.5         &   mas     &      mas/yr          & mas/yr              & km/s        &       mag                                                          &  K&\\

\hline
1 & 1042515801147259008 & (129.799021, 62.501271) & $0.39 \pm 0.03$ & $-33.08 \pm 0.04$ & $-41.03 \pm 0.07$ & $73.9 \pm 1.1\checkmark$&(12.72, 13.26, 12.02)&$4906_{-114}^{+263}$&K&25 \\
2$^\mathrm{M}$ & 1268023196461923712 & (225.783582, 26.246320) & $0.22 \pm 0.02$ & $-29.64 \pm 0.04$ & $-18.88 \pm 0.04$ & $-276.8 \pm 1.6$\checkmark&(13.00, 13.49, 12.35)&$4945_{-80}^{+383}$&K&7 \\
3$^\mathrm{M}$ & 1364548016594914560 & (268.779224, 50.573050) & $0.10 \pm 0.02$ & $-4.39 \pm 0.04$ & $7.82 \pm 0.04$ & $110.4 \pm 0.4$\checkmark&(11.93, 12.56, 11.20)&$4813_{-262}^{+221}$&K&10 \\
4$^\mathrm{B,M}$  & 2106519830479009920 & (285.484415, 45.971657) & $0.12 \pm 0.02$ & $3.30 \pm 0.04$ & $13.17 \pm 0.04$ & $-212.1 \pm 1.0$\checkmark&(12.42, 13.04, 11.69)&$4830_{-162}^{+107}$&K&8 \\
5 & 2233912206910720000 & (299.283801, 55.496959) & $0.28 \pm 0.02$ & $27.85 \pm 0.03$ & $-5.48 \pm 0.03$ & $-343.9 \pm 1.7$\checkmark&(12.97, 13.41, 12.36)&$5158_{-80}^{+802}$&G&11 \\
6$^\mathrm{B, M}$ & 3705761936916676864 & (192.764203, 4.941087) & $0.27 \pm 0.02$ & $15.04 \pm 0.05$ & $-32.29 \pm 0.03$ & $88.7 \pm 1.9$\checkmark&(13.19, 13.66, 12.57)&$5036_{-176}^{+125}$&G&17 \\
7$^\mathrm{M}$ & 3784964943489710592 & (169.356296, -5.815378) & $0.26 \pm 0.04$ & $22.58 \pm 0.08$ & $-16.33 \pm 0.05$ & $126.2 \pm 1.3$\checkmark&(12.25, 12.76, 11.58)&$4997_{-84}^{+174}$&K&9 \\
8 & 4136024785619932800 & (258.736351, -16.502178) & $0.51 \pm 0.09$ & $-1.25 \pm 0.15$ & $-8.23 \pm 0.10$ & $496.5 \pm 4.8$\checkmark&(16.55, 17.04, 15.90)&$4940_{-158}^{+117}$&K&2 \\
9 & 4248140165233284352 & (299.667995, 4.511052) & $0.15 \pm 0.02$ & $-17.34 \pm 0.03$ & $-0.19 \pm 0.03$ & $-358.1 \pm 2.3$\checkmark&(13.21, 13.75, 12.52)&$4859_{-72}^{+89}$&K&7 \\
10 & 4593398670455374592 & (274.896548, 33.818936) & $0.20 \pm 0.02$ & $-1.18 \pm 0.04$ & $-25.74 \pm 0.04$ & $-313.0 \pm 1.2$\checkmark&(12.24, 12.67, 11.65)&$5470_{-442}^{+775}$&K&8 \\
11$^\mathrm{M}$ & 4916199478888664320 & (23.382529, -51.923180) & $0.18 \pm 0.02$ & $-11.09 \pm 0.03$ & $-17.58 \pm 0.04$ & $86.9 \pm 1.3$\checkmark&(12.61, 13.06, 11.99)&$5052_{-69}^{+448}$&G&16 \\
12$^\mathrm{M}$ & 5212817273334550016 & (107.199164, -76.219334) & $0.26 \pm 0.02$ & $12.17 \pm 0.04$ & $35.92 \pm 0.04$ & $159.9 \pm 0.3$\checkmark&(10.89, 11.66, 10.07)&$4245_{-83}^{+160}$&K&8 \\
13 & 5300505902646873088 & (139.033697, -58.890109) & $0.20 \pm 0.01$ & $13.98 \pm 0.03$ & $-16.88 \pm 0.03$ & $160.2 \pm 4.0$\checkmark&(13.19, 13.87, 12.40)& $4363_{-136}^{+90}$&K&3\\ 
14$^\mathrm{M}$ & 5374177064347894272 & (169.498826, -47.831289) & $0.17 \pm 0.02$ & $7.24 \pm 0.04$ & $-17.28 \pm 0.04$ & $143.2 \pm 0.5\checkmark$&(12.19, 12.85, 11.43)&$4761_{-320}^{+106}$&K&17 \\
15 & 5672759960942885376 & (152.033666, -17.673459) & $1.14 \pm 0.05$ & $-4.66 \pm 0.09$ & $6.23 \pm 0.07$ & $-332.5 \pm 2.9$\checkmark&(15.58, 16.07, 14.92)&$4999_{-147}^{+374}$&K&2 \\
16 & 5808433545428565376 & (253.529196, -68.655962) & $0.15 \pm 0.02$ & $-12.04 \pm 0.02$ & $-21.48 \pm 0.02$ & $96.6 \pm 1.0$\checkmark&(13.20, 13.83, 12.46)&$4749_{-164}^{+157}$&K&6 \\
17 & 6053231975369894400 & (181.784844, -64.690105) & $1.31 \pm 0.07$ & $-6.79 \pm 0.12$ & $-2.01 \pm 0.09$ & $-320.1 \pm 2.6$\checkmark&(16.74, 17.45, 15.86)&$4280_{-85}^{+128}$&K&2 \\ 
18$^\mathrm{M}$ & 6397497209236655872 & (333.113416, -68.168596) & $0.17 \pm 0.02$ & $-18.71 \pm 0.02$ & $-6.57 \pm 0.03$ & $-8.2 \pm 3.6$\checkmark&(13.21, 13.68, 12.57)&$5018_{-89}^{+454}$&G&8 \\
19$^\mathrm{B, M}$ & 6431596947468407552 & (274.687922, -70.249323) & $0.08 \pm 0.02$ & $4.55 \pm 0.02$ & $4.97 \pm 0.02$ & $259.1 \pm 1.7$\checkmark&(13.09, 13.66, 12.38)&$4834_{-226}^{+199}$&K&13 \\
20 & 6433337199495213056 & (279.867871, -67.154967) & $0.14 \pm 0.02$ & $-4.15 \pm 0.02$ & $-21.85 \pm 0.02$ & $-89.9 \pm 1.2$\checkmark&(13.00, 13.51, 12.33)&$4893_{-43}^{+74}$&K&14 \\
21 & 6625197335678814208 & (334.068454, -25.560644) & $0.21 \pm 0.03$ & $-7.02 \pm 0.05$ & $-27.30 \pm 0.05$ & $-399.8 \pm 17.8$\checkmark&(13.02, 13.39, 12.46)&$5295_{-334}^{+226}$&G&2 \\
\hline
22 & 1995066395528322560 & (359.273412, 56.883318) & $0.80 \pm 0.03$ & $0.43 \pm 0.04$ & $0.69 \pm 0.04$ & \textcolor{red}{$-799.1 \pm 1.1$}&(13.32, 13.66, 12.81)&$5745_{-388}^{+173}$&G&2 \\ 
23 & 5916830097537967744 & (256.319768, -57.362214) & $0.44 \pm 0.03$ & $-1.05 \pm 0.04$ & $-0.57 \pm 0.03$ & \textcolor{red}{$-457.8 \pm 1.5$}&(13.33, 13.88, 12.57)&$4861_{-208}^{+175}$&K&2 \\
\hline\hline
\end{tabular}}
\label{tab:hvrv1}
In the first column, the superscripts ``B'' and ``M" indicate the sources which are listed in \cite{Bromley2018} and \cite{Marchetti2018}, respectively. The variables are same as Table \ref{tab:rv1}. The numbers shown in red are the \Gaia measurements which are known to be erroneous for reasons given in Section~\ref{sec:spec}, shown ``as is", without being corrected to their true physical values. Their unbound probabilities are shown in Table \ref{tab:hvrv2}.
\end{table*}
\begin{table*}[ht!]
\centering
\scalebox{1}{%
\begin{tabular}{l cccc | cccc|  ccc}
\hline\hline
\GaiaVS & $r_\mathrm{GC}$ & $v_\mathrm{grf}$ & $v_\mathrm{esc}$ & $P_\mathrm{un}$ &    $N_\mathrm{RP}$   & $N_\mathrm{G}$ & GOOD& $RUWE$ & $v_\mathrm{30, esc}$ & $P_\mathrm{30, un}$ &$P_\mathrm{\varpi+0.067, un}$\\
       &            kpc            &   km/s                   &   km/s                     &                              &                                  &                               &            &                 &   km/s                     &                             \\
\hline
1 & $10.285_{-0.170}^{+0.205}$ &$518_{-46}^{+56}$ &$595_{-2}^{+2}$ &0.100 & 0 & 0 &\checkmark& 1.049                           &$595_{-30}^{+30}$ &0.124&0.001\\
2 & $7.753_{-0.057}^{+0.098}$ &$549_{-62}^{+78}$ &$616_{-2}^{+1}$ &0.194 &   0  & 0  &\checkmark& 1.061                         &$616_{-30}^{+30}$ &0.212&0.000\\
3 & $12.029_{-1.177}^{+1.934}$ &$534_{-52}^{+84}$ &$578_{-14}^{+10}$ &0.304 & 0  & 0  &\checkmark& 1.034                     &$576_{-33}^{+32}$ &0.320&0.000\\
4 & $10.190_{-0.588}^{+0.883}$ &$568_{-65}^{+87}$ &$595_{-8}^{+6}$ &0.374 & 0 &0  &\checkmark& 1.024                           &$594_{-31}^{+31}$ &0.386&0.000 \\
5 & $8.958_{-0.081}^{+0.097}$ &$540_{-29}^{+33}$ &$609_{-1}^{+1}$ &0.028 & 0& 0 &\checkmark&0.817                               &$609_{-30}^{+30}$ &0.067&0.000\\
6 & $8.357_{-0.072}^{+0.101}$ &$564_{-48}^{+59}$ &$611_{-2}^{+1}$ &0.215 & 0&0 &\checkmark&0.996                                 &$611_{-30}^{+30}$ &0.243&0.000\\
7 & $9.330_{-0.226}^{+0.342}$ &$530_{-62}^{+86}$ &$602_{-4}^{+3}$ &0.203 & 0 &0 &\checkmark&0.972                                &$602_{-30}^{+30}$ &0.217&0.006\\
8 & $6.390_{-0.380}^{+0.272}$ &$566_{-6}^{+6}$ &$641_{-4}^{+5}$ &0.000 & 0&0 &\checkmark&1.013                                     &$642_{-30}^{+30}$ &0.008&0.000\\
9 & $6.069_{-0.086}^{+0.303}$ &$572_{-62}^{+86}$ &$642_{-5}^{+2}$ &0.212 & 0&0 &\checkmark&0.875                               &$641_{-30}^{+30}$ &0.226&0.000\\
10 & $7.492_{-0.066}^{+0.127}$ &$542_{-58}^{+73}$ &$623_{-2}^{+1}$ &0.141 & 0 &0 &\checkmark&0.907                              &$623_{-30}^{+30}$ &0.159&0.000\\
11 & $9.256_{-0.260}^{+0.374}$ &$533_{-52}^{+68}$ &$600_{-4}^{+3}$ &0.173 & 0 &0 &\checkmark&1.028                              &$600_{-30}^{+30}$ &0.194&0.000 \\
12 & $8.099_{-0.048}^{+0.068}$ &$568_{-50}^{+58}$ &$617_{-1}^{+1}$ &0.202 & 0 & 0 &\checkmark&0.878                              &$617_{-30}^{+30}$ &0.231&0.000 \\
13 & $9.081_{-0.139}^{+0.172}$ &$582_{-34}^{+39}$ &$608_{-2}^{+2}$ &0.259 & 0 & 0 &\checkmark&1.065                               &$608_{-30}^{+30}$ &0.307&0.000\\
14 & $8.682_{-0.283}^{+0.464}$ &$561_{-65}^{+88}$ &$611_{-5}^{+3}$ &0.281 & 0 & 0 &\checkmark&1.023                               &$610_{-30}^{+30}$ &0.296&0.001\\
15 & $8.487_{-0.010}^{+0.011}$ &$564_{-3}^{+3}$ &$615_{-0}^{+0}$ &0.000 & 0 & 0 &\checkmark&0.982                                    &$615_{-30}^{+30}$ &0.046&0.000\\
16 & $5.384_{-0.018}^{+0.082}$ &$573_{-70}^{+87}$ &$651_{-2}^{+1}$ &0.187 & 0 & 0 &\checkmark&0.992                                &$650_{-30}^{+30}$ &0.201&0.000 \\
17 & $7.938_{-0.016}^{+0.015}$ &$546_{-3}^{+3}$ &$622_{-0}^{+0}$ &0.000 & 0 & 0 &\checkmark&1.093                                    &$622_{-30}^{+30}$ &0.006 &0.000\\
18 & $6.825_{-0.056}^{+0.117}$ &$584_{-42}^{+52}$ &$626_{-2}^{+1}$ &0.211 &0 & 0 &\checkmark&1.120                                  &$626_{-30}^{+30}$ &0.245 &0.000 \\
19 & $7.969_{-1.243}^{+2.180}$ &$607_{-59}^{+86}$ &$613_{-21}^{+15}$ &0.475 & 0 & 0 &\checkmark&0.941                              &$611_{-36}^{+34}$ &0.482&0.000 \\
20 & $5.253_{-0.037}^{+0.166}$ &$565_{-69}^{+88}$ &$649_{-3}^{+1}$ &0.178 & 0 & 0 &\checkmark& 0.931                                 &$648_{-30}^{+30}$ &0.191 &0.000\\
21 & $7.149_{-0.028}^{+0.122}$ &$563_{-62}^{+90}$ &$623_{-3}^{+1}$ &0.247 & 0 & 0 &\checkmark&0.957                                   &$622_{-30}^{+30}$ &0.262&0.002\\
\hline
22 & $8.877_{-0.021}^{+0.022}$ &$592_{-1}^{+1}$ &$611_{-0}^{+0}$ &\textcolor{red}{0.000} & 1 & 1 &                    & 1.023       &$611_{-30}^{+30}$ &\textcolor{red}{0.259}&\textcolor{red}{0.000} \\
23 & $6.369_{-0.115}^{+0.103}$ &$604_{-1}^{+1}$ &$642_{-1}^{+2}$ &\textcolor{red}{0.000} & 1 & 1 &                       &1.200       &$642_{-30}^{+30}$ &\textcolor{red}{0.104}&\textcolor{red}{0.000} \\
\hline\hline
\end{tabular}
}
\caption{Unbound probabilities of high velocity star candidates with median radial velocities.
The variables are same as Table \ref{tab:rv2}. 
}
\label{tab:hvrv2}
\end{table*}
\clearpage

\begin{acknowledgements}
We thank David Katz, Douglas Boubert, Nami Mowlavi, Zhengwei Liu, Heran Xiong,  Tommaso Marchetti, Hailiang Chen, Uli Bastian, Jan Rybizki and Haifeng Wang for valuable discussions.
We thank Jianhua Wang and Qi Gao for helping us reduce spectra of \GaiaVS22.
This work has made use of data from the European Space Agency (ESA) mission {\it Gaia} (https://www.cosmos.esa.int/gaia), processed by the {\it Gaia} Data Processing and Analysis Consortium (DPAC, https://www.cosmos.esa.int/web/gaia/dpac/consortium). Funding for the DPAC has been provided by national institutions, in particular the institutions participating in the {\it Gaia} Multilateral Agreement. 
This work is supported by the Natural Science Foundation of China (Nos. 11521303, 11573061, 11733008, 11661161016),
by Yunnan province (Nos. 2015FB190),
by  the Science and Technology Development Fund, Macau SAR (File Nos. 001/2016/AFJ and 0001/2019/A1). This project was developed in part at the 2018 GaiaLAMOST Sprint workshop, supported by the NSFC
under grants 11333003 and 11390372.
\end{acknowledgements}

\bibliographystyle{raa} 
\bibliography{msRAA-2019-0285}

\begin{thebibliography}{73}
\providecommand\natexlab[1]{#1}
\providecommand\JournalTitle[1]{#1}

\bibitem[{Abadi} {et~al.}(2009)]{Abadi2009}
{Abadi}, M.~G., {Navarro}, J.~F., \& {Steinmetz}, M. 2009, \apjl, 691, L63

\bibitem[{Allen} \& {Santillan}(1991)]{Allen1991}
{Allen}, C., \& {Santillan}, A. 1991, \rmxaa, 22, 255

\bibitem[{Arenou} {et~al.}(2018)]{Arenou2018}
{Arenou}, F., {Luri}, X., {Babusiaux}, C., {et~al.} 2018, \aap, 616, A17

\bibitem[{Astraatmadja} \& {Bailer-Jones}(2016)]{Astraatmadja2016}
{Astraatmadja}, T.~L., \& {Bailer-Jones}, C.~A.~L. 2016, \apj, 832, 137

\bibitem[{Blaauw}(1961)]{Blaauw1961}
{Blaauw}, A. 1961, \bain, 15, 265

\bibitem[{Boubert} {et~al.}(2017)]{Boubert2017}
{Boubert}, D., {Erkal}, D., {Evans}, N.~W., \& {Izzard}, R.~G. 2017, \mnras,
  469, 2151

\bibitem[{Boubert} \& {Evans}(2016)]{Boubert2016}
{Boubert}, D., \& {Evans}, N.~W. 2016, \apjl, 825, L6

\bibitem[{Boubert} {et~al.}(2018)]{Boubert2018}
{Boubert}, D., {Guillochon}, J., {Hawkins}, K., {Ginsburg}, I., \& {Evans},
  N.~W. 2018, arXiv:1804.10179

\bibitem[{Boubert} {et~al.}(2019)]{Boubert2019}
{Boubert}, D., {Strader}, J., {Aguado}, D., {et~al.} 2019, \mnras, 486, 2618

\bibitem[{Bromley} {et~al.}(2018)]{Bromley2018}
{Bromley}, B.~C., {Kenyon}, S.~J., {Brown}, W.~R., \& {Geller}, M.~J. 2018,
  \apj, 868, 25

\bibitem[{Brown}(2006)]{Brown2006}
{Brown}, W.~R. 2006, in Bulletin of the American Astronomical Society, Vol.~38,
  American Astronomical Society Meeting Abstracts, 1084

\bibitem[{Brown}(2015)]{Brown2015}
{Brown}, W.~R. 2015, \araa, 53, 15

\bibitem[{Brown} {et~al.}(2009)]{Brown2009}
{Brown}, W.~R., {Geller}, M.~J., \& {Kenyon}, S.~J. 2009, \apj, 690, 1639

\bibitem[{Brown} {et~al.}(2012)]{Brown2012}
{Brown}, W.~R., {Geller}, M.~J., \& {Kenyon}, S.~J. 2012, \apj, 751, 55

\bibitem[{Brown} {et~al.}(2014)]{Brown2014}
{Brown}, W.~R., {Geller}, M.~J., \& {Kenyon}, S.~J. 2014, \apj, 787, 89

\bibitem[{Brown} {et~al.}(2005)]{Brown2005}
{Brown}, W.~R., {Geller}, M.~J., {Kenyon}, S.~J., \& {Kurtz}, M.~J. 2005,
  \apjl, 622, L33

\bibitem[{Brown} {et~al.}(2018)]{Brown2018}
{Brown}, W.~R., {Lattanzi}, M.~G., {Kenyon}, S.~J., \& {Geller}, M.~J. 2018,
  \apj, 866, 39

\bibitem[{Capuzzo-Dolcetta} \& {Fragione}(2015)]{Capuzzo-Dolcetta2015}
{Capuzzo-Dolcetta}, R., \& {Fragione}, G. 2015, \mnras, 454, 2677

\bibitem[Collaboration(2018)]{gaia_dr2light}
Collaboration, G. 2018, Gaia Data Release 2 (DR2) gaia\_source light, {VO}
  resource provided by the {GAVO} Data Center

\bibitem[{Du} {et~al.}(2019)]{Du2019}
{Du}, C., {Li}, H., {Yan}, Y., {et~al.} 2019, \apjs, 244, 4

\bibitem[{Edelmann} {et~al.}(2005)]{Edelmann2005}
{Edelmann}, H., {Napiwotzki}, R., {Heber}, U., {Christlieb}, N., \& {Reimers},
  D. 2005, \apjl, 634, L181

\bibitem[{Erkal} {et~al.}(2019)]{Erkal2019}
{Erkal}, D., {Boubert}, D., {Gualandris}, A., {Evans}, N.~W., \& {Antonini}, F.
  2019, \mnras, 483, 2007

\bibitem[{Evans} {et~al.}(2018)]{Evans2018}
{Evans}, D.~W., {Riello}, M., {De Angeli}, F., {et~al.} 2018, arXiv:1804.09368

\bibitem[{Fritz} {et~al.}(2018)]{Fritz2018}
{Fritz}, T.~K., {Battaglia}, G., {Pawlowski}, M.~S., {et~al.} 2018, \aap, 619,
  A103

\bibitem[{Gaia Collaboration} {et~al.}(2018{\natexlab{a}})]{Gaia2018}
{Gaia Collaboration}, {Brown}, A.~G.~A., {Vallenari}, A., {et~al.}
  2018{\natexlab{a}}, arXiv:1804.09365

\bibitem[{Gaia Collaboration} {et~al.}(2018{\natexlab{b}})]{Helmi2018}
{Gaia Collaboration}, {Helmi}, A., {van Leeuwen}, F., {et~al.}
  2018{\natexlab{b}}, \aap, 616, A12

\bibitem[{Gaia Collaboration} {et~al.}(2018{\natexlab{c}})]{Eyer2018}
{Gaia Collaboration}, {Eyer}, L., {Rimoldini}, L., {et~al.} 2018{\natexlab{c}},
  arXiv:1804.09382

\bibitem[{Geier} {et~al.}(2015)]{Geier2015}
{Geier}, S., {F{\"u}rst}, F., {Ziegerer}, E., {et~al.} 2015, Science, 347, 1126

\bibitem[{Gvaramadze} {et~al.}(2009)]{Gvaramadze2009}
{Gvaramadze}, V.~V., {Gualandris}, A., \& {Portegies Zwart}, S. 2009, \mnras,
  396, 570

\bibitem[{Hills}(1988)]{Hills1988}
{Hills}, J.~G. 1988, \nat, 331, 687

\bibitem[{Hirsch} {et~al.}(2005)]{Hirsch2005}
{Hirsch}, H.~A., {Heber}, U., {O'Toole}, S.~J., \& {Bresolin}, F. 2005, \aap,
  444, L61

\bibitem[{Huang} {et~al.}(2017)]{Huang2017}
{Huang}, Y., {Liu}, X.-W., {Zhang}, H.-W., {et~al.} 2017, \apjl, 847, L9

\bibitem[{Irrgang} {et~al.}(2018)]{Irrgang2018}
{Irrgang}, A., {Kreuzer}, S., \& {Heber}, U. 2018, \aap, 620, A48

\bibitem[{Irrgang} {et~al.}(2013)]{Irrgang2013}
{Irrgang}, A., {Wilcox}, B., {Tucker}, E., \& {Schiefelbein}, L. 2013, \aap,
  549, A137

\bibitem[{Johnson} \& {Soderblom}(1987)]{Johnson1987}
{Johnson}, D.~R.~H., \& {Soderblom}, D.~R. 1987, \aj, 93, 864

\bibitem[{Katz} {et~al.}(2018)]{Katz2018}
{Katz}, D., {Sartoretti}, P., {Cropper}, M., {et~al.} 2018, arXiv:1804.09372

\bibitem[{Kenyon} {et~al.}(2014)]{Kenyon2014}
{Kenyon}, S.~J., {Bromley}, B.~C., {Brown}, W.~R., \& {Geller}, M.~J. 2014,
  \apj, 793, 122

\bibitem[{Kenyon} {et~al.}(2018)]{Kenyon2018}
{Kenyon}, S.~J., {Bromley}, B.~C., {Brown}, W.~R., \& {Geller}, M.~J. 2018,
  \apj, 864, 130

\bibitem[{Kenyon} {et~al.}(2008)]{Kenyon2008}
{Kenyon}, S.~J., {Bromley}, B.~C., {Geller}, M.~J., \& {Brown}, W.~R. 2008,
  \apj, 680, 312

\bibitem[{Koposov} {et~al.}(2019)]{Koposo2019}
{Koposov}, S.~E., {Boubert}, D., {Li}, T.~S., {et~al.} 2019, arXiv e-prints,
  arXiv:1907.11725

\bibitem[{Li} {et~al.}(2015)]{Li2015}
{Li}, Y.-B., {Luo}, A.-L., {Zhao}, G., {et~al.} 2015, Research in Astronomy and
  Astrophysics, 15, 1364

\bibitem[{Li} {et~al.}(2012)]{Li2012}
{Li}, Y., {Luo}, A., {Zhao}, G., {et~al.} 2012, \apjl, 744, L24

\bibitem[{Lindegren} {et~al.}(2018)]{Lindegren2018}
{Lindegren}, L., {Hern{\'a}ndez}, J., {Bombrun}, A., {et~al.} 2018, \aap, 616,
  A2

\bibitem[{Lu} {et~al.}(2010)]{Luyoujun2010}
{Lu}, Y., {Zhang}, F., \& {Yu}, Q. 2010, \apj, 709, 1356

\bibitem[{Marchetti} {et~al.}(2018{\natexlab{a}})]{Marchetti2018a}
{Marchetti}, T., {Contigiani}, O., {Rossi}, E.~M., {et~al.} 2018{\natexlab{a}},
  \mnras, 476, 4697

\bibitem[{Marchetti} {et~al.}(2018{\natexlab{b}})]{Marchetti2018}
{Marchetti}, T., {Rossi}, E.~M., \& {Brown}, A.~G.~A. 2018{\natexlab{b}},
  \mnras, 2466

\bibitem[{Navarro} {et~al.}(1997)]{Navarro1997}
{Navarro}, J.~F., {Frenk}, C.~S., \& {White}, S.~D.~M. 1997, \apj, 490, 493

\bibitem[{Odenkirchen} \& {Brosche}(1992)]{Odenkirchen1992}
{Odenkirchen}, M., \& {Brosche}, P. 1992, Astronomische Nachrichten, 313, 69

\bibitem[{Palladino} {et~al.}(2014)]{Palladino2014}
{Palladino}, L.~E., {Schlesinger}, K.~J., {Holley-Bockelmann}, K., {et~al.}
  2014, \apj, 780, 7

\bibitem[{Pauli} {et~al.}(2003)]{Pauli2003}
{Pauli}, E.-M., {Napiwotzki}, R., {Altmann}, M., {et~al.} 2003, \aap, 400, 877

\bibitem[{Pauli} {et~al.}(2006)]{Pauli2006}
{Pauli}, E.-M., {Napiwotzki}, R., {Heber}, U., {Altmann}, M., \& {Odenkirchen},
  M. 2006, \aap, 447, 173

\bibitem[{Raddi} {et~al.}(2018{\natexlab{a}})]{Raddi2018b}
{Raddi}, R., {Hollands}, M.~A., {Gaensicke}, B.~T., {et~al.}
  2018{\natexlab{a}}, arXiv:1804.09677

\bibitem[{Raddi} {et~al.}(2018{\natexlab{b}})]{Raddi2018a}
{Raddi}, R., {Hollands}, M.~A., {Koester}, D., {et~al.} 2018{\natexlab{b}},
  \apj, 858, 3

\bibitem[{Raddi} {et~al.}(2019)]{Raddi2019}
{Raddi}, R., {Hollands}, M.~A., {Koester}, D., {et~al.} 2019, \mnras, 489, 1489

\bibitem[{Sch{\"o}nrich}(2012)]{Schonich2012}
{Sch{\"o}nrich}, R. 2012, \mnras, 427, 274

\bibitem[{Sch{\"o}nrich} {et~al.}(2010)]{Schonich2010}
{Sch{\"o}nrich}, R., {Binney}, J., \& {Dehnen}, W. 2010, \mnras, 403, 1829

\bibitem[{Shen} {et~al.}(2018)]{Shen2018}
{Shen}, K.~J., {Boubert}, D., {G{\"a}nsicke}, B.~T., {et~al.} 2018,
  arXiv:1804.11163

\bibitem[{Sohn} {et~al.}(2018)]{Sohn2018}
{Sohn}, S.~T., {Watkins}, L.~L., {Fardal}, M.~A., {et~al.} 2018, \apj, 862, 52

\bibitem[{Tauris}(2015)]{Tauris2015}
{Tauris}, T.~M. 2015, \mnras, 448, L6

\bibitem[{Tauris} \& {Takens}(1998)]{Tauris1998}
{Tauris}, T.~M., \& {Takens}, R.~J. 1998, \aap, 330, 1047

\bibitem[{Taylor}(2005)]{Topcat}
{Taylor}, M.~B. 2005, in Astronomical Society of the Pacific Conference Series,
  Vol. 347, Astronomical Data Analysis Software and Systems XIV, ed.
  P.~{Shopbell}, M.~{Britton}, \& R.~{Ebert}, 29

\bibitem[{Tillich} {et~al.}(2009)]{Tillich2009}
{Tillich}, A., {Przybilla}, N., {Scholz}, R.-D., \& {Heber}, U. 2009, \aap,
  507, L37

\bibitem[{Vennes} {et~al.}(2017)]{Vennes2017}
{Vennes}, S., {Nemeth}, P., {Kawka}, A., {et~al.} 2017, Science, 357, 680

\bibitem[{Wang} \& {Han}(2009)]{Wang2009}
{Wang}, B., \& {Han}, Z. 2009, \aap, 508, L27

\bibitem[{Wang} {et~al.}(2013)]{Wang2013}
{Wang}, B., {Justham}, S., \& {Han}, Z. 2013, \aap, 559, A94

\bibitem[{Watkins} {et~al.}(2018)]{Watkins2018}
{Watkins}, L.~L., {van der Marel}, R.~P., {Sohn}, S.~T., \& {Evans}, N.~W.
  2018, arXiv:1804.11348

\bibitem[{Wilkinson} \& {Evans}(1999)]{Wilkinson1999}
{Wilkinson}, M.~I., \& {Evans}, N.~W. 1999, \mnras, 310, 645

\bibitem[{Yu} \& {Tremaine}(2003)]{Yu2003}
{Yu}, Q., \& {Tremaine}, S. 2003, \apj, 599, 1129

\bibitem[{Zhang} {et~al.}(2010)]{Zhangfupeng2010}
{Zhang}, F., {Lu}, Y., \& {Yu}, Q. 2010, \apj, 722, 1744

\bibitem[{Zhao} {et~al.}(2018)]{Zhao2018}
{Zhao}, Y., {Fan}, Z., {Ren}, J.-J., {et~al.} 2018, Research in Astronomy and
  Astrophysics, 18, 110

\bibitem[{Zheng} {et~al.}(2014)]{Zheng2014}
{Zheng}, Z., {Carlin}, J.~L., {Beers}, T.~C., {et~al.} 2014, \apjl, 785, L23

\bibitem[{Zhong} {et~al.}(2014)]{Zhong2014}
{Zhong}, J., {Chen}, L., {Liu}, C., {et~al.} 2014, \apjl, 789, L2

\bibitem[{Ziegerer} {et~al.}(2015)]{Ziegerer2015}
{Ziegerer}, E., {Volkert}, M., {Heber}, U., {et~al.} 2015, \aap, 576, L14

\end{thebibliography}

\begin{appendix}
\renewcommand{\thesection}{\Alph{section}}

\section{1}
\label{sec:grtv}

The Galactic rest frame velocity can be expressed as
\begin{equation}
\mathbf{v}_\mathrm{grf} = \mathbf{R}\cdot\mathbf{v} + \mathbf{v}_{\odot},
\end{equation}
\noindent where $\mathbf{R}=\mathbf{T}\cdot\mathbf{A}$, in which $\mathbf{T}$ is the rotation matrix from equatorial coordinates to Galactic coordinates, and $\mathbf{A}$ is the coordinate matrix of $\mathbf{v}$ (details in \citealt{Johnson1987}, the J2000 rotation matrix to Galactic coordinates is taken from the introduction to the Hipparcos catalog); $\mathbf{v} = (v_\mathrm{s},  \frac{k\mu_{\alpha}}{\varpi}, \frac{k\mu_{\delta}}{\varpi})^T$, where $k = 4.740470446$ km/s, and $\mathbf{v}_{\odot}$ is the Solar velocity in the Galactic rest frame. It follows that

\begin{equation}
v_\mathrm{grf}^2 = av^2_\mathrm{s} + bv_\mathrm{s} +c,
\end{equation}
where
\begin{equation}
\begin{array}{l}
a = R^2_{11} + R^2_{21} + R^2_{31},\\
b = 2 (R_{11} A + R_{21} B+ R_{31} C),\\
c = A^2 + B^2 +C^2,
\end{array}
\end{equation}
and
\begin{equation}
\begin{array}{l}
A = R_{12} \frac{k\mu_{\alpha}}{\varpi} + R_{13} \frac{k\mu_{\delta}}{\varpi}+ v_{\odot1}, \\
B = R_{22} \frac{k\mu_{\alpha}}{\varpi} + R_{23} \frac{k\mu_{\delta}}{\varpi}+ v_{\odot2},\\
C = R_{32} \frac{k\mu_{\alpha}}{\varpi} + R_{33} \frac{k\mu_{\delta}}{\varpi}+ v_{\odot3}.
\end{array}
\end{equation}

\end{appendix}

\end{document}